\documentclass[]{aa}

\usepackage{graphicx,epsf,psfig,epsfig,times,amsmath,amssymb}

\usepackage{natbib}
\bibpunct{(}{)}{;}{a}{}{,}

\def \BB {{\sc BB}}
\def \compps {{\sc compps}}
\def \comptt {{\sc comptt}}
\def \PL {{\sc pl}}
\def \COMPBB {{\sc compbb}}
\def \DISKBB {{\sc diskbb}}
\def \BMC {{\sc BMC}}
\def \XSPEC {{\sc XSPEC}}

\begin{document}

\title{Average hard X-ray emission from NS LMXBs: Observational evidence of different spectral states in
NS LMXBs\thanks{Based on observations with \textit{INTEGRAL}, an ESA project
with instruments and science data centre funded by ESA member states
(especially the PI countries: Denmark, France, Germany, Italy,
Spain, and Switzerland), Czech Republic and Poland, and with the
participation of Russia and the USA.}
}
  \author{A. Paizis\inst{1}, R. Farinelli\inst{2}, L. Titarchuk\inst{2,}\inst{3,}\inst{4}, T.J.-L.
   Courvoisier\inst{5,}\inst{6}, 
    A. Bazzano\inst{7}, V. Beckmann\inst{3,}\inst{8}, F. Frontera\inst{2}, P. Goldoni\inst{9,}\inst{10}, E.
    Kuulkers\inst{11}, S. Mereghetti\inst{1},
     J. Rodriguez\inst{10}, O. Vilhu\inst{12} }
\offprints{A. Paizis, ada@iasf-milano.inaf.it}
\institute{
INAF-IASF, Sezione di Milano, Via Bassini 15, I--20133 Milano, Italy
\and Dipartimento di Fisica, Universit\`{a} di Ferrara, Via Saragat 1, I--44100 Ferrara, Italy 
\and NASA Goddard Space Flight Center, Exploration of the Universe Division, Greenbelt, MD 20771, USA
\and George Mason University/Center for Earth Observing and Space Research, Fairfax, VA 22030; 
and US Naval Research Laboratory, Code 7655, Washington, DC 20375-5352
\and \textit{INTEGRAL} Science Data Centre, Chemin d'Ecogia 16, 1290 Versoix, Switzerland
\and Observatoire de Gen\`eve, 51 chemin des Mailletes, CH--1290 Sauverny, Switzerland
\and INAF-IASF, Sezione di Roma, Via del Fosso del Cavaliere 100, I-00133, Roma, Italy
\and Joint Center for Astrophysics, Department of Physics, University of Maryland, Baltimore County, MD 21250, USA
\and APC/UMR 7164, 11 Place M. Berthelot,75231 Paris, France
\and CEA, Centre de Saclay, DSM/DAPNIA/SAp, 91191 Gif-sur-Yvette Cedex France
\and ISOC, ESA/ESAC, Urb. Villafranca del Castillo, P.O. Box 50727, 28080, Madrid, Spain
\and Observatory, P.O.Box 14, T\"ahtitorninm\"aki, Fi--00014 University of Helsinki, Finland
}
  \date{Received 12 June 2006 / Accepted 10 August 2006}

   \abstract {}
   {We studied and compared the long-term average hard X-ray ($>$20\,keV)
   spectra of a sample of twelve bright low-mass X-ray binaries hosting a neutron star (NS).
   Our sample comprises the six well studied Galactic
 Z sources and six Atoll sources,  four of which are bright ("GX")
bulge sources while two are weaker ones in the 2--10\,keV range (H~1750--440 and H~1608--55). }
   { For all the sources of our sample, we analysed
   available public data
    and extracted average spectra from the IBIS/ISGRI detector on board \textit{INTEGRAL}.  }
   { We can describe all the spectral states 
   in terms of the bulk motion Comptonisation scenario. We find evidence that
   bulk motion 
   is always present, its strength is related to the accretion rate and it is suppressed only in the
   presence of high local luminosity.
The two low-dim Atoll source spectra  are dominated by   photons up-scattered
presumably  due to dynamical and thermal Comptonisation in  an optically thin, hot plasma.
For the first time, we extend the detection of H~1750--440 up to 150\,keV.
   The Z  and bright "GX" Atoll source spectra are very similar and are
   dominated by Comptonised blackbody radiation of seed photons, presumably coming from
   the accretion disc and NS surface, in an
   optically thick cloud with plasma temperature in the range of 2.5--3\,keV.
Six sources show a  hard tail in their  \emph{average} spectrum:
Cyg~X-2 (Z), GX~340$+$0 (Z), GX~17$+$2 (Z), GX~5--1 (Z), Sco~X--1 (Z) and GX~13$+$1 (Atoll).
This is the first detection of a hard tail
in the X-ray spectrum of the peculiar GX~13$+$1.
 Using radio data from the literature
we find, in all Z sources and bright "GX" Atolls, a \emph{systematic}
positive correlation between the X-ray hard tail (40--100\,keV) and the radio luminosity.
This  suggests that hard tails and energetic electrons  causing  the radio
emission may have the same origin,  most likely the Compton cloud
located inside the NS magnetosphere. }
   {}
\authorrunning{Paizis et al.}
\titlerunning{Average hard X-ray emission from NS LMXBs}
\maketitle

%

\section{Introduction}

Low-Mass X-ray Binaries (LMXBs) are systems where a compact object, either 
a neutron star (NS) or a black hole candidate (BHC), accretes matter via Roche lobe
overflow from a companion with a mass  \mbox{$\textit{M}\lesssim$ $1\textit{M}_{\odot}$}.
NS LMXBs can be broadly classified 
according to their timing and spectral properties
\citep{hasinger89}. On the basis of this classification, NS LMXBs are divided in 
Z sources and Atoll sources from the shape of their track in the colour-colour diagram
and from the different timing behaviour that correlates with the position on 
the tracks.

The overall spectra of Z sources are very soft 
 \citep[][and references therein]{barret02} and can be
described by the sum of a cool ($\sim$1\,keV) blackbody (BB) and a Comptonised emission
from an electron plasma ("corona") of a few keV.
Instead, Atoll sources 
perform quite dramatic spectral changes: when bright, they can have soft spectra 
(similar to Z sources) but they  switch to  low/hard spectra at low luminosities.
\cite{titarchuk05}, hereafter TS05, implemented a thorough analysis of spectral 
and temporal properties of the Atoll source 4U~1728--34. They show that  the  low/hard spectra at low 
luminosities  can be described by the sum of  up-scattered    spectra related to the Comptonisation 
of the disc and NS surface  soft photons. 
They found that  the Compton cloud electron temperature is  of the order of  a few tens of keV.
These spectra are very similar to the hard-state spectra of BHCs but they 
are softer (BHC photon index $\Gamma \sim 1.6\pm 0.1$ vs NS $\Gamma\sim 2.1\pm 0.1$) 
as expected from the theory \citep[see, for example,][]{titarchuk04}.
TS05 found that the high luminosity state spectrum of 4U~1728--34 consists of the sum of two pure 
blackbody-like spectra with colour temperatures of about 1\,keV and 2.2\,keV. The softer BB component 
is presumably related to disc emission as the harder one is related to the NS emission.
It is worth noting that TS05 also found that  when the source undergoes hard-soft transition, all
 power spectrum (PDS) frequencies (QPO and break frequencies) increase with the photon index, 
 with no sign of saturation. Note that the index-QPO correlation observed in BHC shows 
 the index saturation at high values of QPO frequency \citep[see][]{shaposhnikov06}.
 
So far, only Atoll sources (and more generally X-ray bursters) have been observed with 
low/hard spectra (i.e. Comptonising corona of few \emph{tens} of keV).
 Z sources always have soft Comptonisation spectra (Comptonising corona of few keV) and 
 can have an 
 additional hard X-ray component dominating the spectrum above
$\sim$30\,keV. This component is  on top of the soft spectrum and is 
highly variable with  most of the emission remaining soft \citep[see][for a review on NS LMXB spectra]{barret01, disalvo02}.
 Hence, we would expect that the average high energy spectra 
of Atoll sources have a strong component above 30\,keV and that in the soft Z sources 
this component is less prominent and smeared out 
in the time averaged spectrum.  
Z sources spend most of their time in the high/soft state, but   
 they may show the transition to  harder states at lower luminosity.  
The sensitivity of the past missions may have introduced an observational bias, 
similarly to the lack of a continuous
coverage of the Galactic plane and Centre in the less explored hard X-ray range (above 20\,keV). 
Moreover,  the concentration of these sources towards the Galactic centre makes it difficult to 
observe them with
non-imaging instruments. Consequently, data analysis and interpretation of such observations 
is extremely problematic.
\begin{figure*}
\centering
\includegraphics[width=1.0\linewidth]{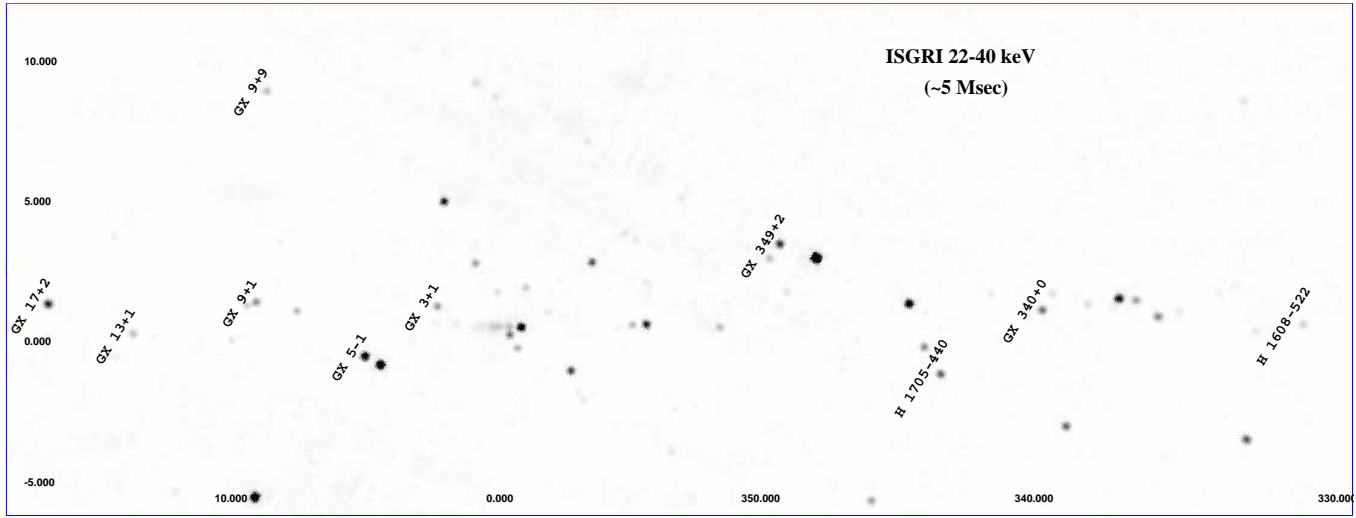}
\caption{IBIS/ISGRI 22--40\,keV mosaic of the Galactic bulge (about 5\,Msec).
Only the sources studied in this paper are labeled. The location of Cyg~X--2 and Sco~X--1 
is not covered by this map.
\label{fig:ima1}}
\end{figure*}
All these instrumental biases can be minimised with the use of the recently launched 
INTErnational Gamma-Ray Astrophysics Laboratory, \textit{INTEGRAL} 
\citep{winkler03}.
The imager  \textit{INTEGRAL}/IBIS \citep{ubertini03} has high sensitivity, 
about $\sim$10 times better than \textit{GRANAT}/SIGMA, coupled to imaging capability with 12$^{\prime}$ angular 
resolution above 20\,keV. 

In this paper we report the study of the average hard X-ray spectra of twelve NS LMXBs 
 performed with the low energy (20--200\,keV) IBIS detector, ISGRI \citep{lebrun03},
 using a coherent and large sample of data, free
from systematic effects which play a role when combining data from different missions.
 The sample of the LMXBs chosen is given in Table~\ref{tab:table1} and comprises  six
 Galactic Z sources and six Atoll sources,  four of which are bright ("GX") 
bulge sources while two are weaker ones in the 2-10\,keV range.

Our approach is two-fold: on one side, for comparison purposes, we study the average spectra in terms of 
phenomenological models as done in the literature, on the other,
we study the sources in the frame of a physical model in the attempt to find  a 
self-consistent  scenario that describes all the spectral properties we observe. We discuss 
the similarities of such a scenario with the  black hole LMXB case as well as the radio - 
X-ray correlation
that is typical of LMXBs.

\begin{table}
  \begin{center}

    \caption{LMXBs studied in this paper.  \emph{D}: distance (in kpc) from references in \cite{migliari06} except for (*)  from \cite{christian97}
    and (**) from \cite{ford00};
    \emph{Rate}: average \mbox{22--40\,keV} counts/sec of the source 
    as obtained from the mosaic image shown 
    in Fig.~\ref{fig:ima1}. Multiply by $\sim$10 to obtain a flux estimate in units of mCrab. 
    F(1mCrab)$_{22-40\,keV}$ $\sim$6.8$\times$10$^{-12}$\rm\,erg\,s$^{-1}$\,cm$^{-2}$;
    \emph{SNR}: signal to noise ratio 
    in the 22--40\,keV band;  \emph{MaxEn}: maximum energy channel (keV)
    with a signal to noise ratio higher than three in the average spectrum; \emph{T$_{exp}$}: 
    effective exposure time in ksec.
         }\vspace{1em}
    \renewcommand{\arraystretch}{1.2}
    \begin{tabular}[h]{cccccc}
      \hline
Source &  D &  Rate          & SNR &  MaxEn & T$_{exp}$\\
	&(kpc) & (cps)  &         &     (keV)	& (ksec)	\\	
\hline
\hline
Z sources &	\\
\hline
Sco~X--1&  2.8 & 58.3 &  1757 &  150 & 266\\
GX~340$+$0  & 11& 2.4 &  124 & 46 & 433\\
GX~349$+$2   & 5& 3.4&  189 &  39 & 265\\ 
GX~5--1   & 9.2 & 3.8 &330  &  80 & 1091\\
GX~17$+$2 &14 & 4.4& 203 & 80 & 248\\
Cyg~X--2    & 13.3& 2.2   & 57& 55& 149\\
\hline
Atoll sources &\\
\hline
H~1608--522  & 4** &0.8 &36 &   150 & 299\\
H~1705--440 &11** & 3 &144 & 150 & 307\\
GX~9$+$9& 5* &0.97 &  55 &  35 & 139\\
GX~3$+$1& 5.6*&  0.96 & 86 &  43 &2027\\
GX~9$+$1& 7* &1.2&91 & 37 & 491\\
GX~13$+$1 & 7 &0.96 &  57 & 80 & 290\\
\hline
      \end{tabular}
    \label{tab:table1}
  \end{center}
\end{table}

\section{Observations and data analysis}

We have analysed  \textit{INTEGRAL} data publically available in  which the sources in
Table~\ref{tab:table1} were in the Fully Coded Field of View (FCFOV) of IBIS.
The additional criterion of a minimum of good time of 1000\,sec for IBIS/ISGRI 
led to a total of 2263 pointings  each with variable exposure time  
(from about 1800\,sec up to about 3600\,sec) spanning from January 2003 to May 2004. 
Version 5.1 of the Off-line Scientific Analysis (OSA) software
has been used to analyse the data. The description of the algorithms used in the 
IBIS/ISGRI scientific analysis can be found in  \cite{goldwurm03}. \\
For each pointing we extracted images in the \mbox{22--40} and \mbox{40--80\,keV} energy bands. 
The images were used to build 
 light-curves as well as a final mosaic. The mosaic 
images   revealed the sources that 
were active in the field of view at the time of the observations. The list of all the detected
sources (one per source of interest) was then used in the spectral step, where 
spectra for all the active sources in the field of view were simultaneously 
extracted with the standard spectral extraction method.
The single pointing spectra were then averaged into one final spectrum per source. 
These final average spectra have been used in our spectral 
study adding  1\% systematic error (with the exception of the bright Sco~X--1 for which we 
added 1.5\% systematic error). In the fit, only data points with more than 3$\sigma$ 
in the \mbox{22--200\,keV} range have been considered.\\
To cross-check our results, we also used  the alternative method of extracting spectra
from the mosaics. To do so, we re-ran the imaging step in twelve different bins 
(instead of the  22--40 and 40--80\,keV previously mentioned) for all the 
pointings and then extracted a spectrum using the
flux from each energy map. We verified that the two different methods
 give compatible results. In this work we show only 
the results of the former method (the standard extraction) for which we have a finer binning.

\begin{figure*}
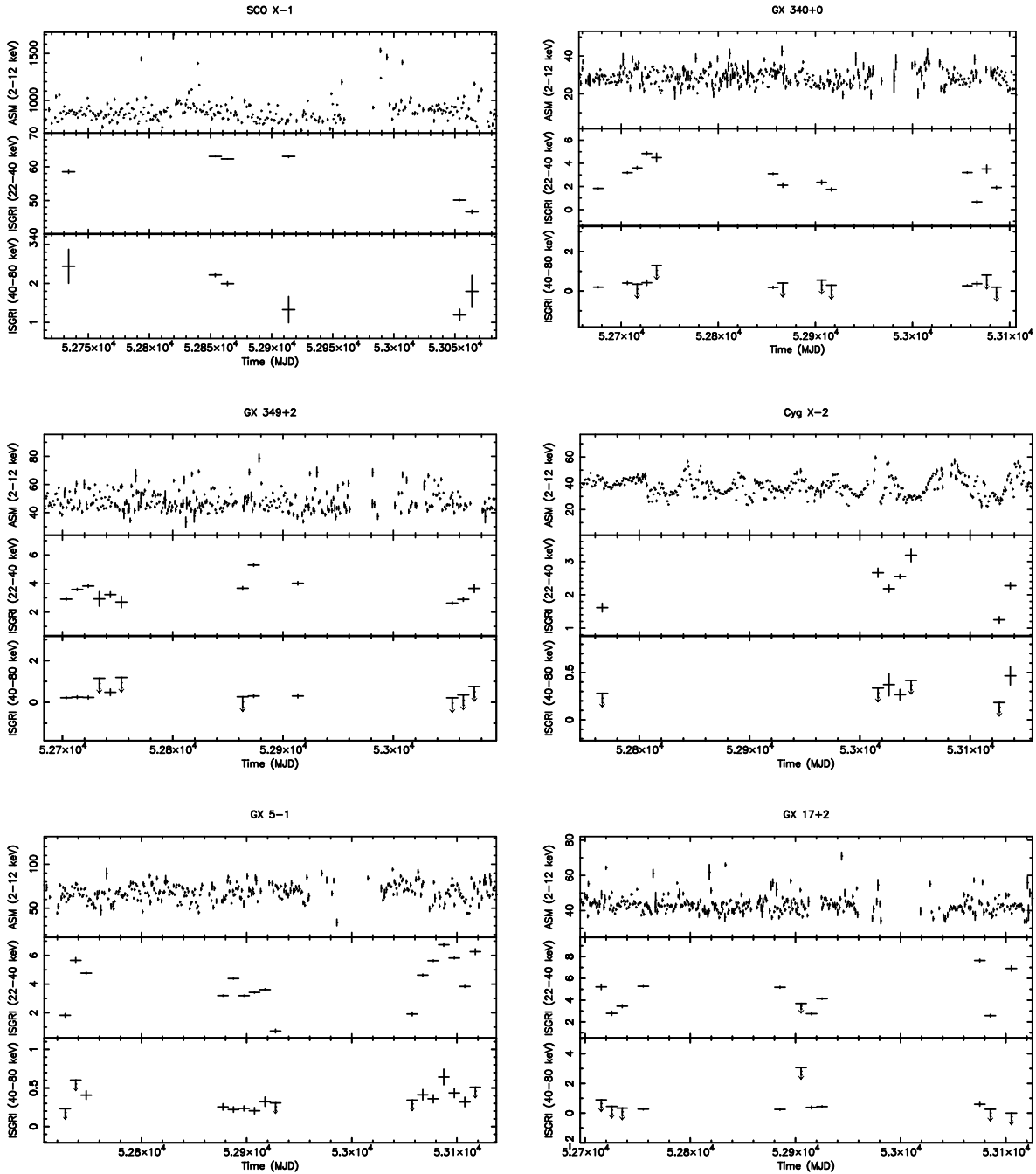


\hbox{\hspace{0.5cm}
\includegraphics[height=7.5cm,angle=-90]{./5792f2.ps}
\hspace{0.5cm}
\includegraphics[height=7.5cm,angle=-90]{./5792f3.ps}}
\vbox{\vspace{0.15cm}}

\hbox{\hspace{0.5cm}
\includegraphics[height=7.5cm,angle=-90]{./5792f4.ps}
\hspace{0.5cm}
\includegraphics[height=7.5cm,angle=-90]{./5792f5.ps}}
\vbox{\vspace{0.15cm}}

\hbox{\hspace{0.5cm}
\includegraphics[height=7.5cm,angle=-90]{./5792f6.ps}
\hspace{0.5cm}
\includegraphics[height=7.5cm,angle=-90]{./5792f7.ps}}

\caption[]{\textit{RXTE}/ASM and IBIS/ISGRI light-curves (counts/sec) of the Z sources. 
One Crab is approximately 
75\,counts/sec in \textit{RXTE}/ASM \mbox{(2--12\,keV)}, 93\,counts/sec in IBIS/ISGRI 
22--40\,keV and 61\,counts/sec in IBIS/ISGRI 40--80\,keV. }
\label{fig:Zlc}
\end{figure*}

\begin{figure*}

\hbox{\hspace{0.5cm}
\includegraphics[height=7.5cm,angle=-90]{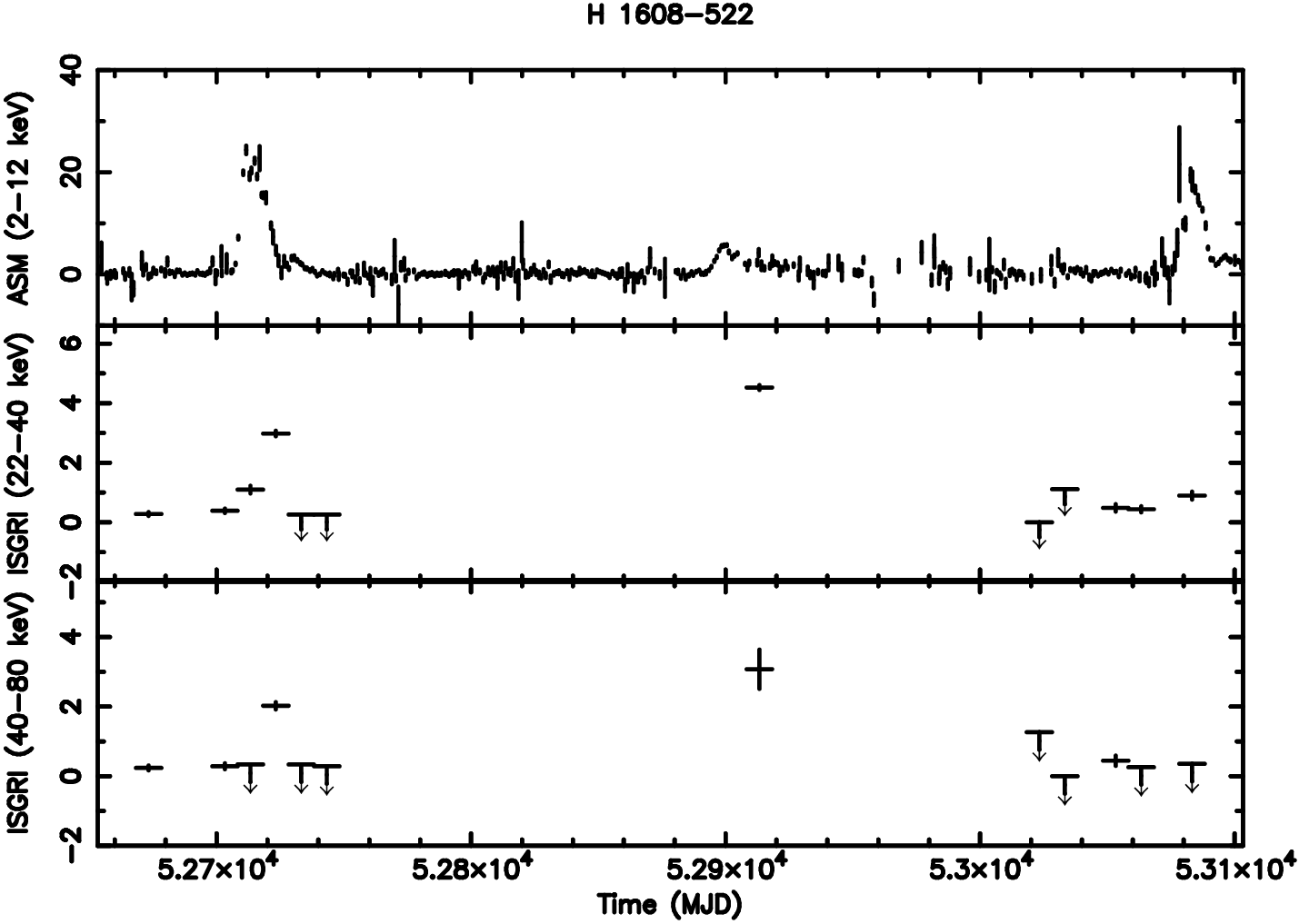}
\hspace{0.5cm}
\includegraphics[height=7.5cm,angle=-90]{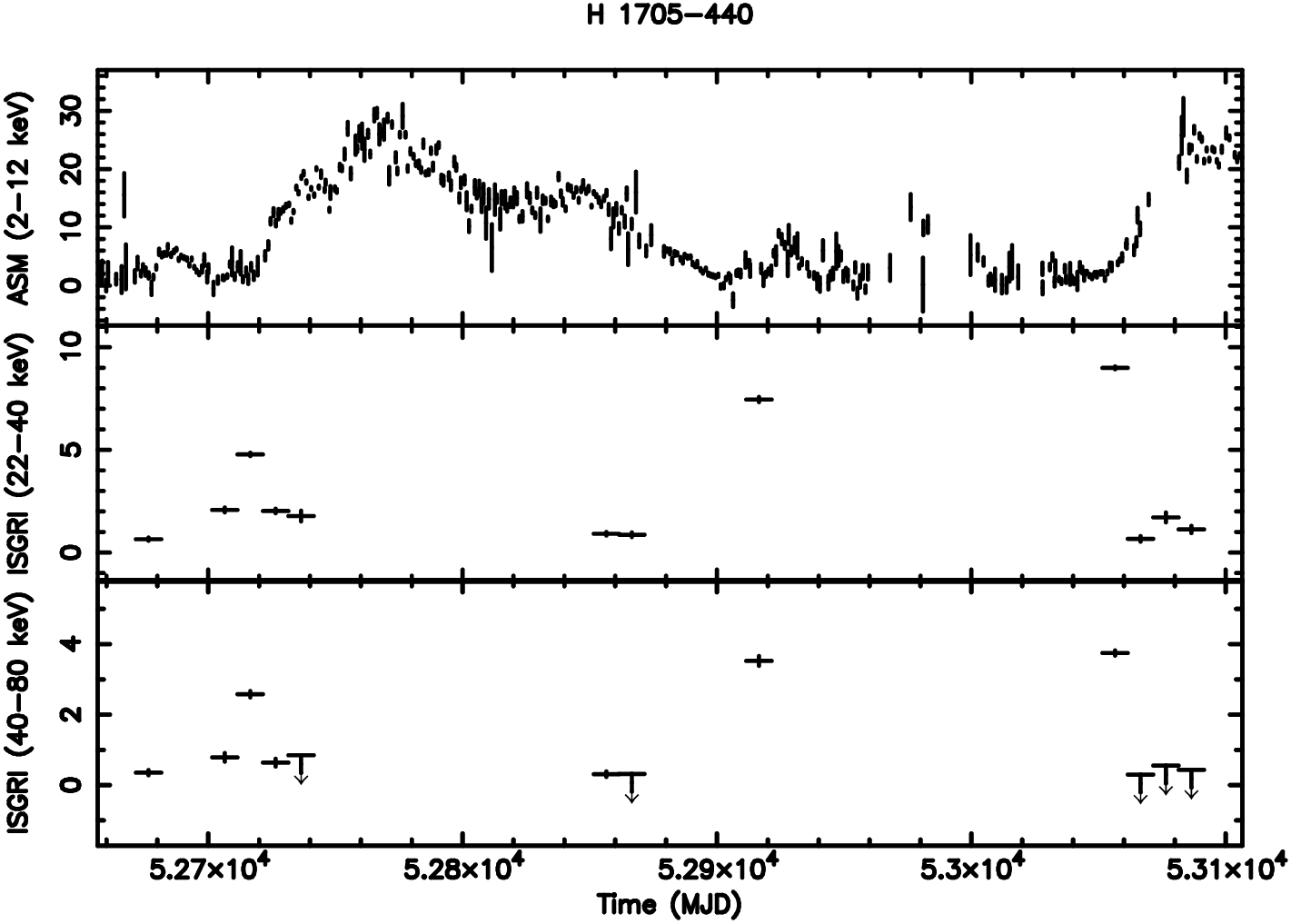}}
\vbox{\vspace{0.15cm}}

\hbox{\hspace{0.5cm}
\includegraphics[height=7.5cm,angle=-90]{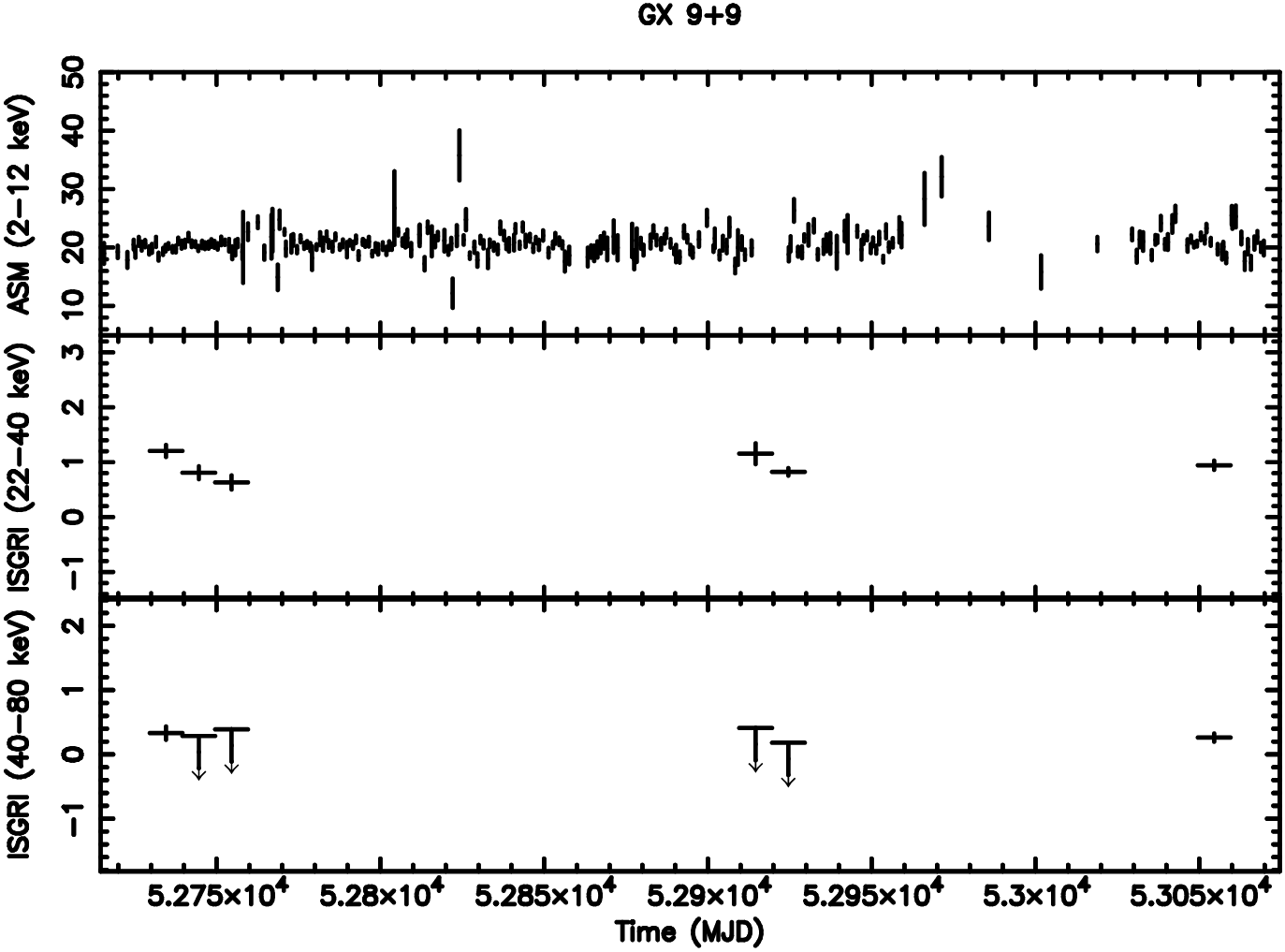}
\hspace{0.5cm}
\includegraphics[height=7.5cm,angle=-90]{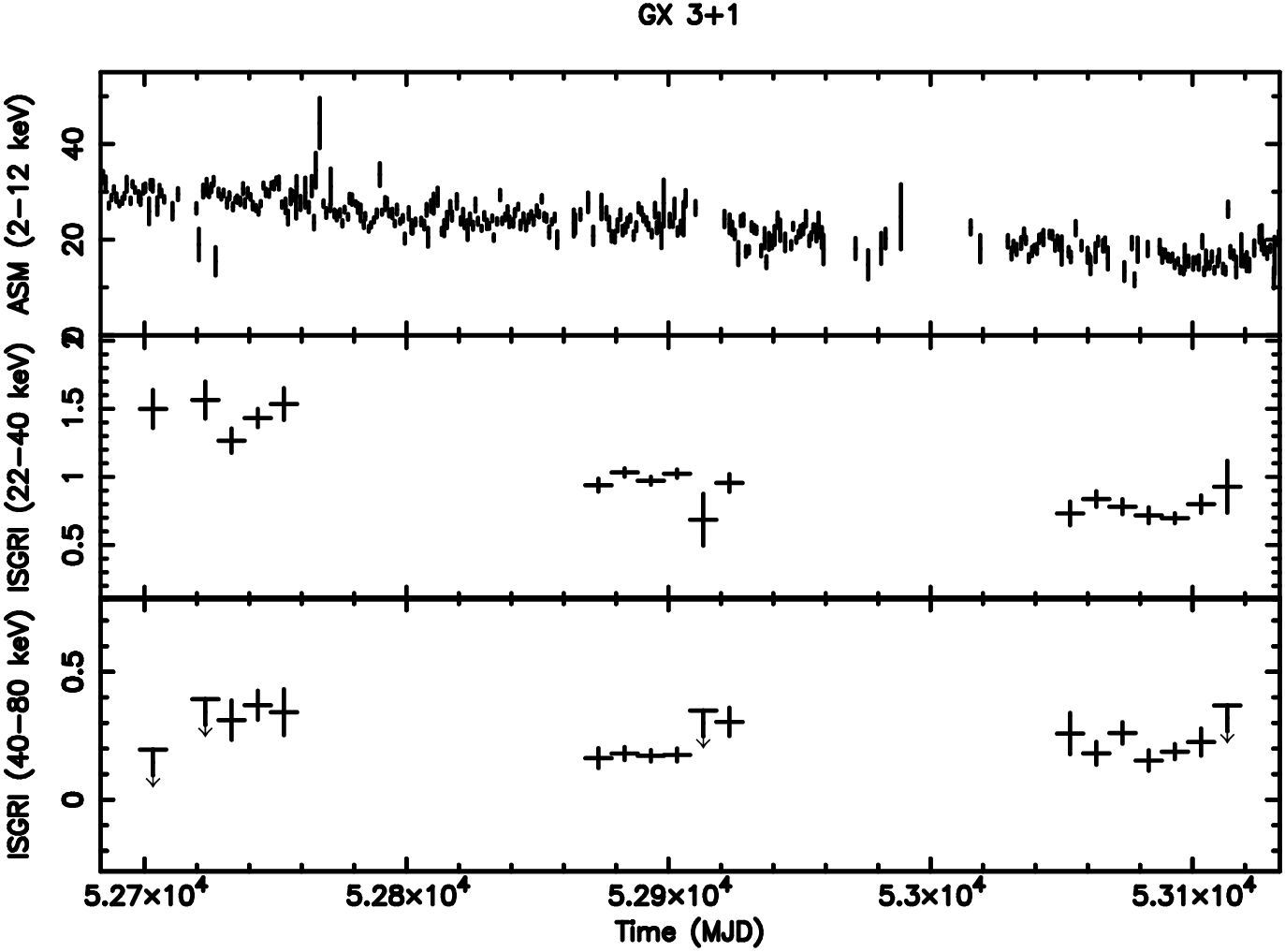}}
\vbox{\vspace{0.15cm}}

\hbox{\hspace{0.5cm}
\includegraphics[height=7.5cm,angle=-90]{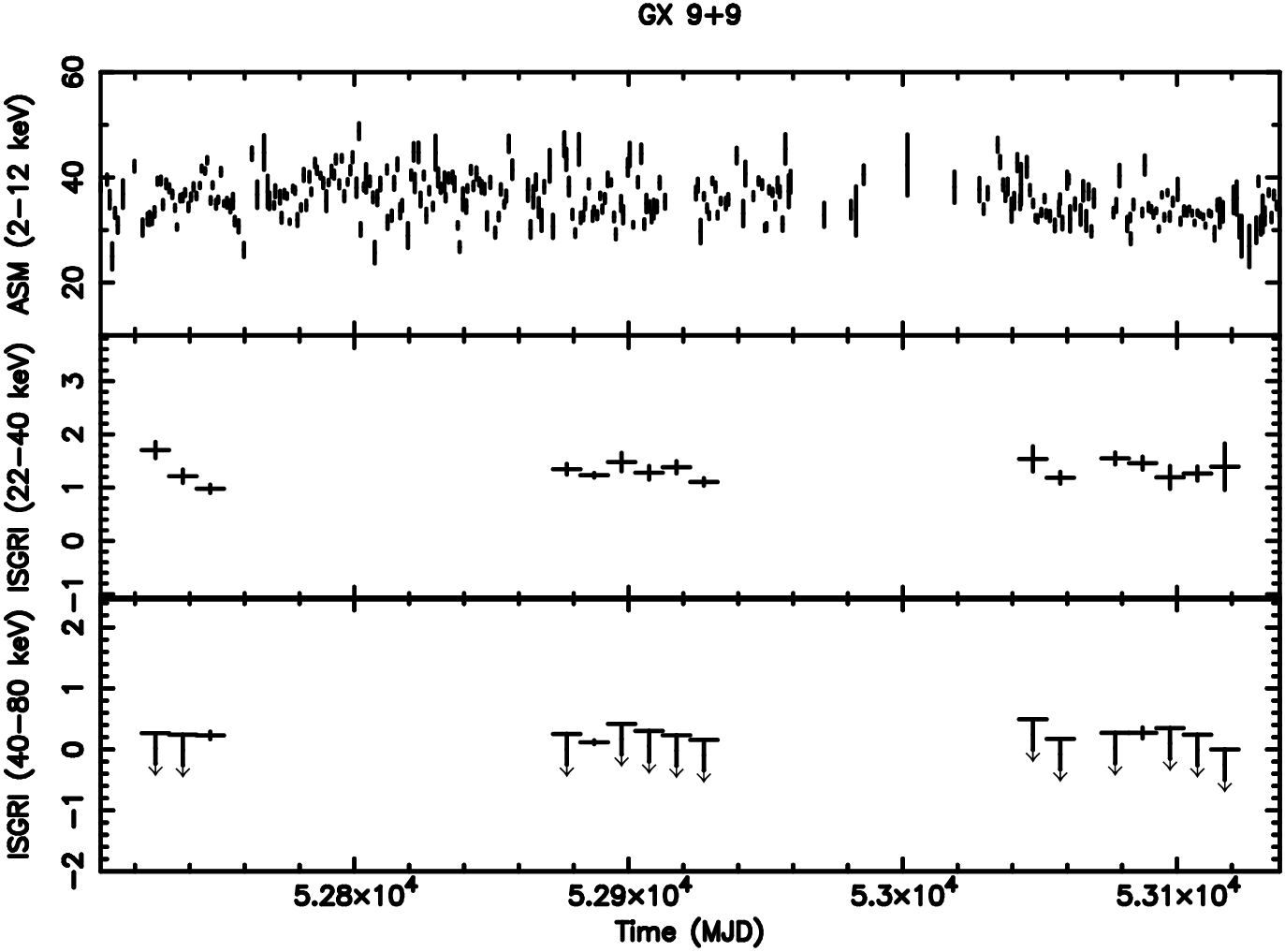}
\hspace{0.5cm}
\includegraphics[height=7.5cm,angle=-90]{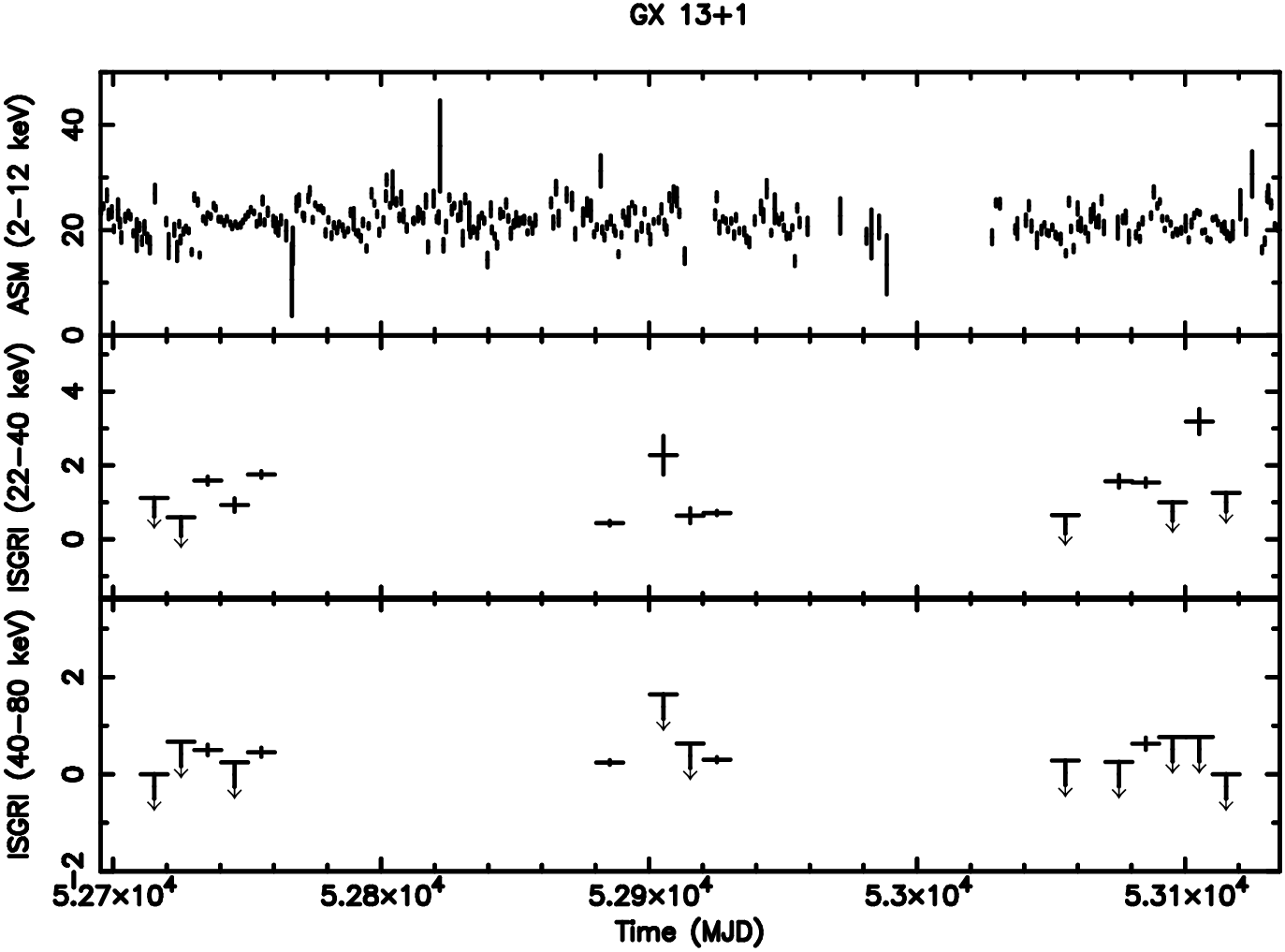}}

\caption[]{\textit{RXTE}/ASM and IBIS/ISGRI light-curves (counts/sec) of the Atoll sources. 
One Crab is approximately 
75\,counts/sec in \textit{RXTE}/ASM \mbox{(2--12\,keV)}, 93\,counts/sec in IBIS/ISGRI 
22--40\,keV and 61\,counts/sec in IBIS/ISGRI 40--80\,keV. }
\label{fig:Alc}
\end{figure*}

\section{Results}
 
Figure~\ref{fig:ima1} shows the final sky map
of the pointings around the Galactic bulge in the 22--40\,keV band. 
The sources studied in this paper are labeled, with the exception of 
\mbox{Cyg~X--2} and \mbox{Sco~X--1} that lie outside 
the image. 
The 22--40\,keV average count rate and detection significance of each source, 
taken from the mosaic image 
shown in Fig.~\ref{fig:ima1}, 
are reported in Table~\ref{tab:table1}. The  count rate is in  counts/sec and 
multiplied  by $\sim$10 gives an estimate in mCrab\footnote{The
    average Crab count rate  in 
    different positions of the 
    FCFOV in the same band is about 93\,counts/sec.}.

Figures~\ref{fig:Zlc} and~\ref{fig:Alc} show the IBIS/ISGRI (\mbox{22--40\,keV} and \mbox{40--80\,keV}) 
 light-curves as obtained from the imaging analysis (re-binned to ten day data bins). The simultaneous
quick-look results provided by the \textit{RXTE}/ASM team are also shown (one day data bins). One Crab is approximately 
75\,counts/sec in the \textit{RXTE}/ASM 2--12\,keV energy band, 93\,counts/sec and 61\,counts/sec in the IBIS/ISGRI 22--40 and 40--80\,keV 
band, respectively.

 \subsection{A phenomenological approach}

As a first attempt we tried to fit all the spectra with the same simple model, in order to compare them directly. 
It soon turned out that  this was not possible since the long-term average spectra 
identified three main spectral states. 
Phenomenologically, we can classify them in terms of a widely used 
thermal Comptonisation model, \comptt~by \cite{titarchuk94}. We found that 
 the spectra of  GX~349$+$2, GX~9$+$1, 
GX~9$+$9, and GX~3$+$1 are well described by  a single \comptt~component while we need two components,  
\comptt+\PL~(thermal Comptonisation plus power-law), to describe the data  
for  \mbox{Sco~X--1}, \mbox{GX~5--1}, \mbox{GX~340$+$0}, \mbox{GX~17$+$2}, \mbox{Cyg~X--2} and \mbox{GX~13$+$1}. 
On the other hand,  
the spectral shapes  of  \mbox{H~1608--522} and \mbox{H~1705--440} can be well fitted by  a simple
power-law, \PL. 

\begin{figure}
\centering
\includegraphics[angle=270, width=0.9\linewidth]{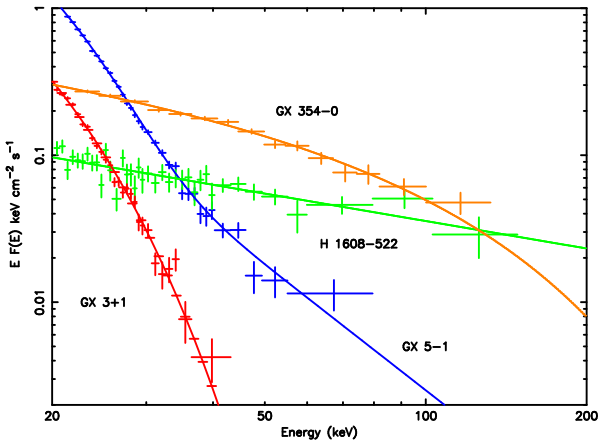}
\caption{IBIS/ISGRI average spectra of GX~354--0 ("pure" low/hard state, see text), H~1608--522 (low/hard state), 
GX~5--1 (intermediate state) and GX~3$+$1 (very soft state).
\label{fig:comp}}
\end{figure}

\begin{figure*}
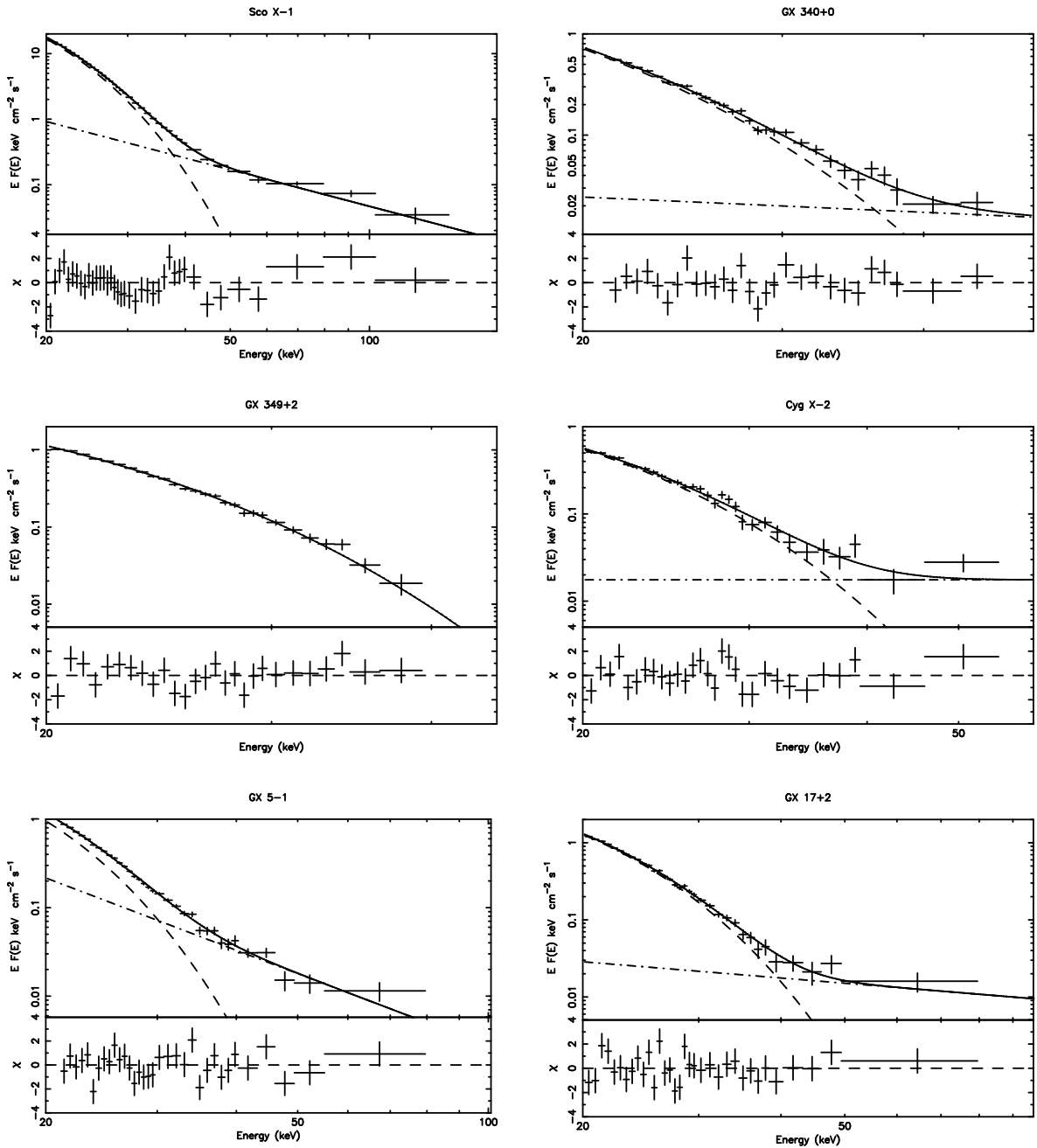


\hbox{\hspace{0.5cm}
\includegraphics[height=7.5cm,angle=-90]{5792f15.ps}
\hspace{0.5cm}
\includegraphics[height=7.5cm,angle=-90]{5792f16.ps}}
\vbox{\vspace{0.15cm}}

\hbox{\hspace{0.5cm}
\includegraphics[height=7.5cm,angle=-90]{5792f17.ps}
\hspace{0.5cm}
\includegraphics[height=7.5cm,angle=-90]{5792f18.ps}}
\vbox{\vspace{0.15cm}}

\hbox{\hspace{0.5cm}
\includegraphics[height=7.5cm,angle=-90]{5792f19.ps}
\hspace{0.5cm}
\includegraphics[height=7.5cm,angle=-90]{5792f20.ps}}

\caption[]{Z sources: average IBIS/ISGRI spectra and best fit models shown in Table~\ref{tab:fit}.}
\label{fig:Zspe}
\end{figure*}

By comparison with  black hole LMXBs, hereafter we call the  
spectral states fitted with a single \comptt~a \emph{very soft} state 
 for which  the spectral shape is well described by the blackbody-like 
 shape slightly modified by Comptonisation.  
Using this analogy, we call our  \comptt+\PL-state  an  \emph{intermediate} state. 
In fact,  the  \emph{intermediate} state in the  BHCs is characterized by the blackbody-like component  at low energies and the 
 steep power-law component at higher energies.
 In the low/hard state spectra  of 
BHCs, the blackbody signature is smeared out but the power-law is prominent. Keeping in mind  this observational 
 fact for the BHCs,  we can call  our  \PL~state  a \emph{low/hard} state.
The three states are shown in Fig.~\ref{fig:comp}. In the plot we have also included the spectrum of the Atoll source 
\mbox{GX~354--0} \citep{falanga06} to point out that the  \emph{low/hard} state of our classification, associated to
\mbox{H~1608--522} in the plot, is not a "pure" low/hard state as in the case of \mbox{GX~354--0} (thermal Comptonisation with cut-off) 
but is a low/hard state where an additional physical mechanism producing a non-attenuated power-law starts to be important
 (see Sect. 4.1). 
Nevertheless, for simplicity hereafter  we will 
 refer to the spectra of  \mbox{H~1608-522} and \mbox{H~1705--440} as to \emph{low/hard} states.
Table~\ref{tab:fit} summarises the spectral models used to fit the data while the corresponding plots with the 
spectra and best fit models are shown in Figs.~\ref{fig:Zspe} and~\ref{fig:ASPE}. 

\begin{table*}
  \begin{center}
    \caption{Summary of the spectral properties of the sources.  No error means the parameter was fixed to the 
    indicated value.  \emph{Model}: best fit model. See text for details; 
    \emph{$\Gamma$}: photon index of the power-law; \emph{kT$_{seed}$}: temperature of seed photons for 
    Comptonisation in keV; \emph{kT$_{e}$}: Comptonising plasma temperature in keV; 
    \emph{$\tau$}: optical depth of the Comptonising plasma; \emph{Flux$_{(20-40\,keV)}$}: flux obtained between 20--40\,keV. 
    For the sources detected up to
    150\,keV also the 40--150\,keV flux is given.}\vspace{1em}
    \renewcommand{\arraystretch}{1.2}
    \begin{tabular}[h]{ccccccccc}
      \hline
Source &  Model & $\Gamma$ & kT$_{seed}$ & kT$_{e}$ & $\tau$ &  $\chi^{2}$/dof & Flux$_{(20-40\,keV)}$ &  Flux$_{(40-150\,keV)}$  \\
       &      &       &   (keV)     & (keV)    &        &                 & (\rm\,erg\,s$^{-1}$\,cm$^{-2}$) &(\rm\,erg\,s$^{-1}$\,cm$^{-2}$)\\                          
      \hline
      \hline
Sco~X$-$1&\comptt+\PL  & $3.85{+0.01 \atop -0.08}$& [1.3] & $2.87{+0.01 \atop -0.02}$& $9.45{+0.11 \atop -0.34}$& 42/34 & 5.8$\times$10$^{-9}$ & 2.2$\times$10$^{-10}$\\

GX~340$+$0 &  \comptt+\PL& [2.5] & [0.93]& [3.0] & 7.73$^{+   0.93}_{-   0.78}$& 24/26 &  2.4$\times$10$^{-10}$ & \\

GX~349$+$2 & \comptt & - & [1.3] & $2.75{+0.33 \atop -0.15}$ & $12{+16 \atop -5}$ &22/24 &  3.5$\times$10$^{-10}$ & \\ 

GX~5--1  & \comptt+\PL & $4.7{+0.6 \atop -0.11}$& [1.0]& $2.42{+0.41 \atop -0.06}$& $25{+5 \atop -16}$& 35/29 &   3.5$\times$10$^{-10}$  & \\

GX~17$+$2 &  \comptt+\PL & [2.7] &[0.6] &$2.94$$\pm$0.04 & [11] & 35/31 & 4.5$\times$10$^{-10}$ &\\

Cyg~X--2  &  \comptt+\PL& [1.96]& [1.09]&$3.02{+0.10 \atop -0.10}$ & 10.68 & 29/27 & 2.1$\times$10$^{-10}$ \\
\hline
H~1608--522 &\PL & 2.62 $\pm$ 0.11 & - & - & - & 45/36 &  8.7$\times$10$^{-11}$ &  9.1$\times$10$^{-11}$ \\
       	     & \compps & - & 2.29${+1.14 \atop -0.25}$ & $122{+24 \atop -43}$ & $0.18{+0.43 \atop -0.13}$ &40/34 \\

H~1705--440 &  \PL & 3.15 $\pm$ 0.04 & - & - & -  & 36/37 & 3$\times$10$^{-10}$ & 1.9$\times$10$^{-10}$\\
	    &  \compps & - & $2.56{+0.62 \atop -0.47}$ & 47 $\pm$ 7 & $0.63{+0.18 \atop -0.17}$ & 31/35 \\

GX~9$+$9&  \comptt& - & [1.3] & $3.66{+25 \atop -1.00}$ & 6($>$0.4) & 27/23 &  9.2$\times$10$^{-11}$ \\

GX~3$+$1& \comptt & - & [1.3] & $2.88{+0.59 \atop -0.28}$& $7.5{+4.7 \atop -2.9}$ & 46/24 &  9.3$\times$10$^{-11}$ \\

GX~9$+$1&   \comptt & -& [1.3]& $2.30{+0.45 \atop -0.08}$ & 26 ($>$ 7)& 32/22 &  1.3$\times$10$^{-10}$\\

GX~13$+$1 &  \comptt+\PL & [2.8] & [1.0]  & $2.75{+0.25 \atop -0.23}$ & [8] & 23/22 & 8.2$\times$10$^{-11}$\\

\hline
      \end{tabular}
    \label{tab:fit}
  \end{center}
\end{table*}

\begin{table*}
  \begin{center}

    \caption{Best fit parameters for the bulk motion Comptonisation model (BMC). 
    No error means the parameter was fixed to the 
    indicated value.  \emph{kT$_{bb}$}: BB colour temperature, in keV;
    \emph{$\alpha$}: energy spectral index ($\Gamma$=$\alpha$ $+$1);  \emph{$\log {A}$}: covering 
    of the BB by the Compton cloud  ; \emph{Flux$_{(20-40\,keV)}$}: flux obtained between 20--40\,keV. 
    For the sources detected up to
    150\,keV also the 40--150\,keV flux is given.
         }\vspace{1em}
    \renewcommand{\arraystretch}{1.2}
    \begin{tabular}[h]{ccccccc}
      \hline
Source &   kT$_{bb}$&    $\alpha$        & $\log {A}$ &   $\chi^{2}$/dof & Flux$_{(20-40\,keV)}$ & Flux$_{(40-150\,keV)}$    \\
       &   (keV)   &                    &           &                  &  (\rm\,erg\,s$^{-1}$\,cm$^{-2}$) &(\rm\,erg\,s$^{-1}$\,cm$^{-2}$)\\                          
\hline
\hline
Sco~X--1&  2.56$^{+   0.01}_{-   0.01}$ & 3.27$^{+   0.06}_{-   0.02}$ & -1.47$^{+   0.01}_{-   0.01}$ &  68/35 & 5.8$\times$10$^{-9}$ &2.2$\times$10$^{-10}$ \\

GX~340$+$0  & 2.43$^{+   0.18}_{-   0.31}$& $<$6 & -1.12$^{+   0.81}_{-   1.05}$ & 24/25& 2.45$\times$10$^{-10}$ &\\

GX~349$+$2   &  2.47$^{+   0.04}_{-   0.12}$ & $<$6   & -1.62$^{+   0.69}_{-   1.03}$   &20/23  &3.5$\times$10$^{-10}$\\

GX~5--1   & 2.35$^{+   0.07}_{-   0.09}$ & 3.84$^{+   0.72}_{-   0.66}$ & -1.06$^{+   0.28}_{-   0.26}$  &34/30  & 3.75$\times$10$^{-10}$ &\\

GX~17$+$2  & 2.68$^{+   0.06}_{-   0.06}$  & 2.0$^{+   1.46}_{-   1.41}$ & -1.87$^{+   0.55}_{-   0.42}$ &32/30 & 4.5$\times$10$^{-10}$ &\\

Cyg~X--2    & 2.78$^{+   0.08}_{-   0.07}$  & $<$1.8  & -0.56$^{+   0.11}_{-   0.14}$  &26/26 &  2.1$\times$10$^{-10}$ &\\

\hline

H~1608--522  & 1.49$^{+   1.39}_{-   0.73}$ &  1.50$^{+   0.14}_{-   0.16}$  & -1.24$^{+   1.45}_{-   2.5}$  & 39/34 & 8.5$\times$10$^{-11}$ & 9.6$\times$10$^{-11}$\\

H~1705--440 & 0.84$^{+   0.14}_{-   0.84}$  & 2.15$^{+   0.05}_{-   0.05}$ & -2.13$^{+  10.13}_{-   3.85}$   & 36/35 &  2.9$\times$10$^{-10}$ & 1.9$\times$10$^{-10}$\\

GX~9$+$9&  2.53$^{+   0.18}_{-   0.17}$ &  $<$6  & -0.70$^{+   0.34}_{-   1.36}$  &  24/22& 9.4$\times$10$^{-11}$ \\

GX~3$+$1& 2.30$^{+   0.11}_{-   0.14}$  & $<$6 & -1.27$^{+   0.37}_{-   1.32}$ & 42/23 & 9.3$\times$10$^{-11}$ & \\

GX~9$+$1& 2.20$^{+   0.05}_{-   0.07}$   & $<$3.2 & -1.09$^{+   0.16}_{-   0.33}$   & 24/21 &1.32$\times$10$^{-10}$\\

GX~13$+$1 &2.32$^{+   0.27}_{-   0.36}$  & 2.02$^{+   1.48}_{-   1.33}$  & -1.34$^{+   0.58}_{-   0.50}$ & 23/23 & 8.2$\times$10$^{-11}$ & \\
\hline
      \end{tabular}
    \label{tab:bmc}
  \end{center}
\end{table*}

\subsubsection{The very soft spectral state} \label{sec:comptt}
Three bright Atoll sources (GX~3$+$1, GX~9$+$1 and GX~9$+$9)  and the Z source GX~349$+$2  
can be well described by a simple \comptt~component. A hard tail has been observed to dominate the 
spectra above 30\,keV in a pointed observation of \textbf{GX~349$+$2} \citep{disalvo01}, but 
it is likely not frequent/bright enough to be detected in a 265\,ksec average spectrum. 
Indeed, no hard tail was detected in the spectrum of GX~349$+$2 by \cite{iaria04} 
using \textit{BeppoSAX} data even though the source was in the same position in the Z-track where 
the tail was first discovered.
No hard tails have ever been detected in the spectra of GX~9$+$9, GX~9$+$1 and GX~3$+$1.

Only recent results have appeared as far as the X-ray spectrum of \textbf{GX~9$+$9} is concerned.
\cite{kong06}, based on \textit{RXTE} data, report a spectrum that can be described by the typical two component 
model used for bright LMXBs hosting a neutron star: a softer component 
(e.g. blackbody) plus a Comptonised harder one (e.g. \comptt).
We are not sensitive to the soft blackbody component with IBIS/ISGRI and 
our results as far as the \comptt~part are concerned are in agreement with what was found by 
\cite{kong06} i.e. a Comptonising plasma kT$_{e}$  of about 4\,keV with a high optical depth $\tau$ about 7. 
Our \textit{RXTE-INTEGRAL} simultaneous observations \citep{vilhu06} show that the broad band 
spectrum of GX~9$+$9 can be also well described by the sum of a soft component (\DISKBB~in \XSPEC~terminology) 
and thermal Comptonisation (\COMPBB) by a hot  optically thin electron plasma.

\begin{figure*}

\hbox{\hspace{0.5cm}
\includegraphics[height=7.5cm,angle=-90]{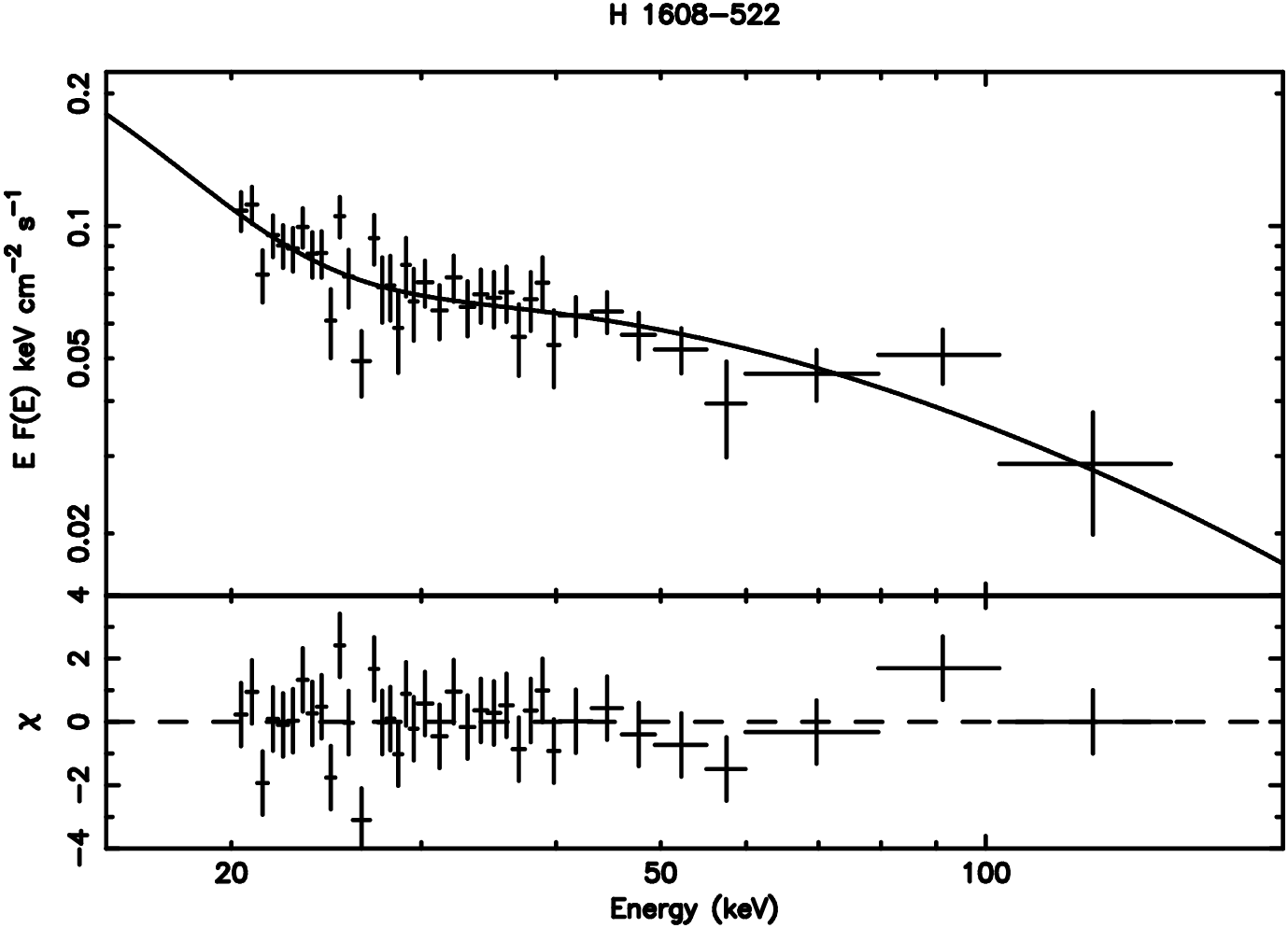}
\hspace{0.5cm}
\includegraphics[height=7.5cm,angle=-90]{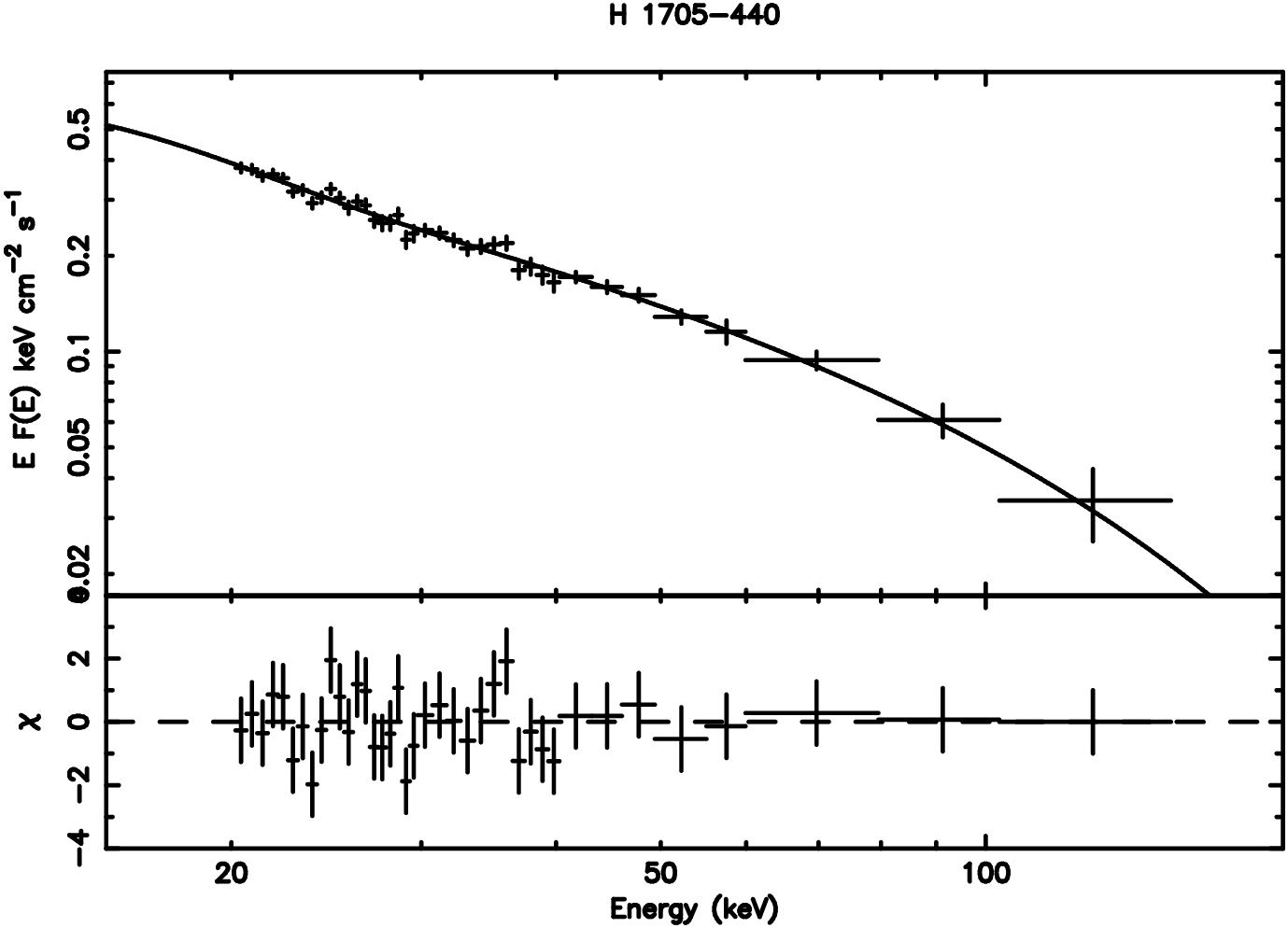}}
\vbox{\vspace{0.15cm}}

\hbox{\hspace{0.5cm}
\includegraphics[height=7.5cm,angle=-90]{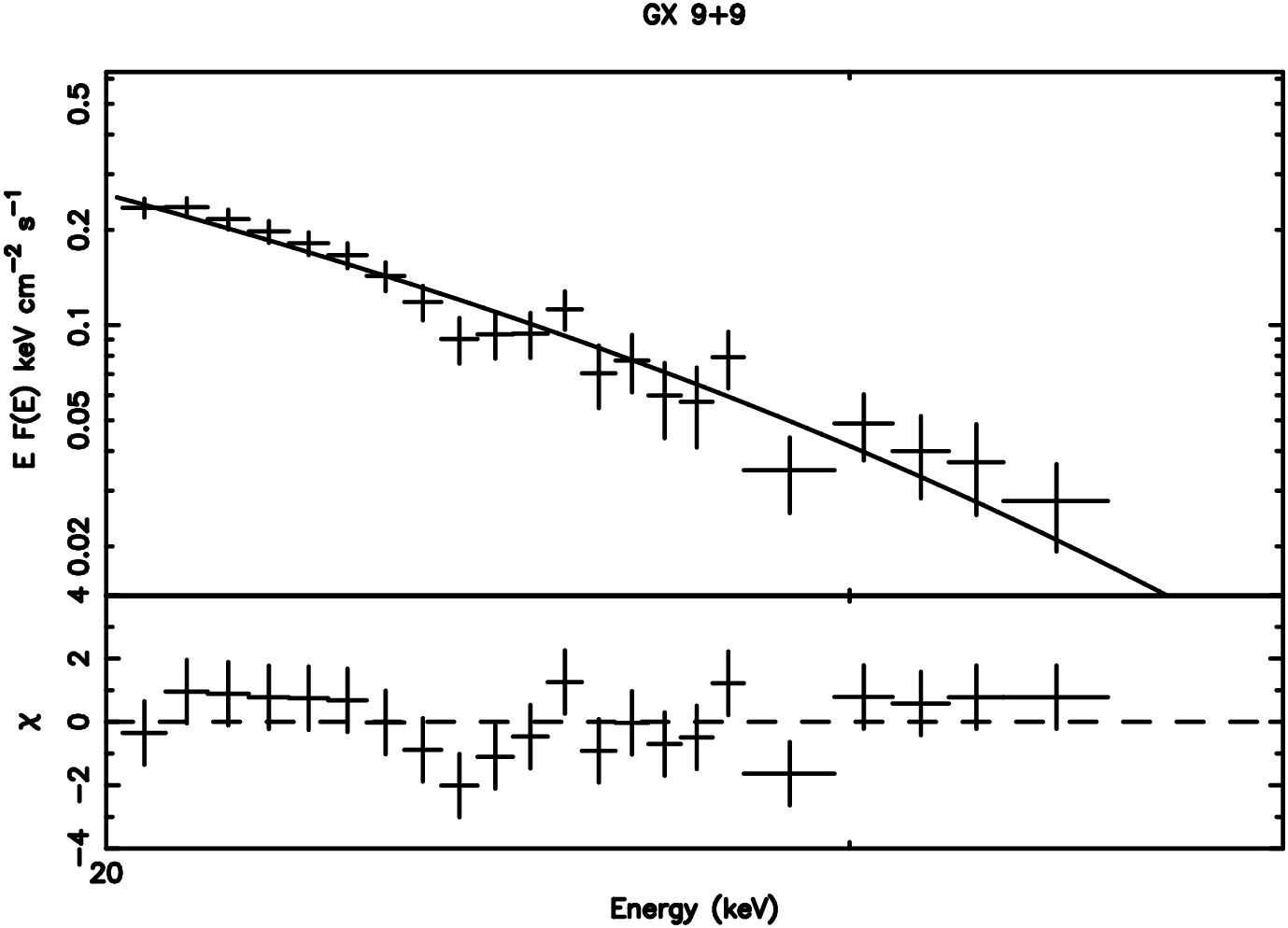}
\hspace{0.5cm}
\includegraphics[height=7.5cm,angle=-90]{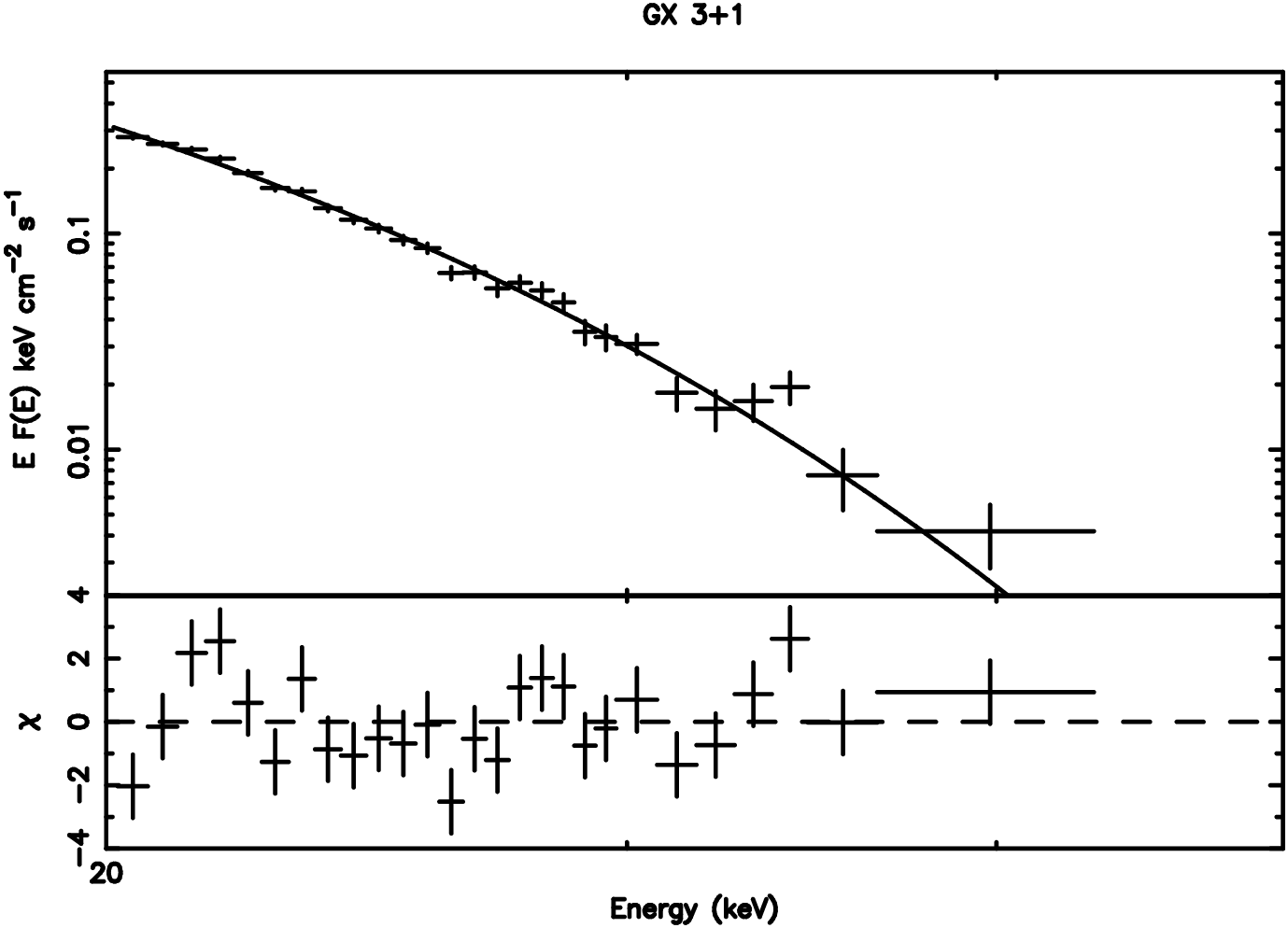}}
\vbox{\vspace{0.15cm}}

\hbox{\hspace{0.5cm}
\includegraphics[height=7.5cm,angle=-90]{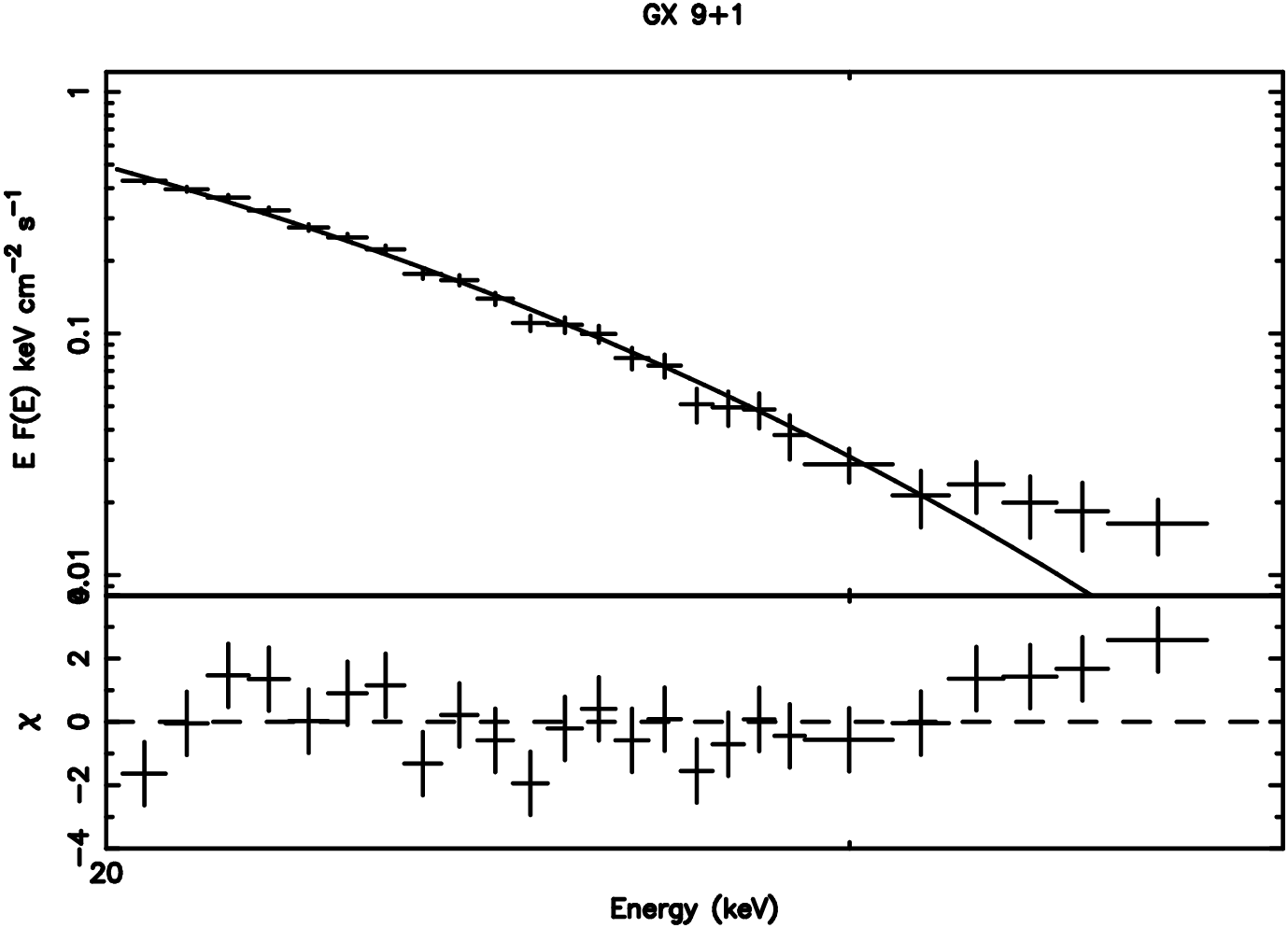}
\hspace{0.5cm}
\includegraphics[height=7.5cm,angle=-90]{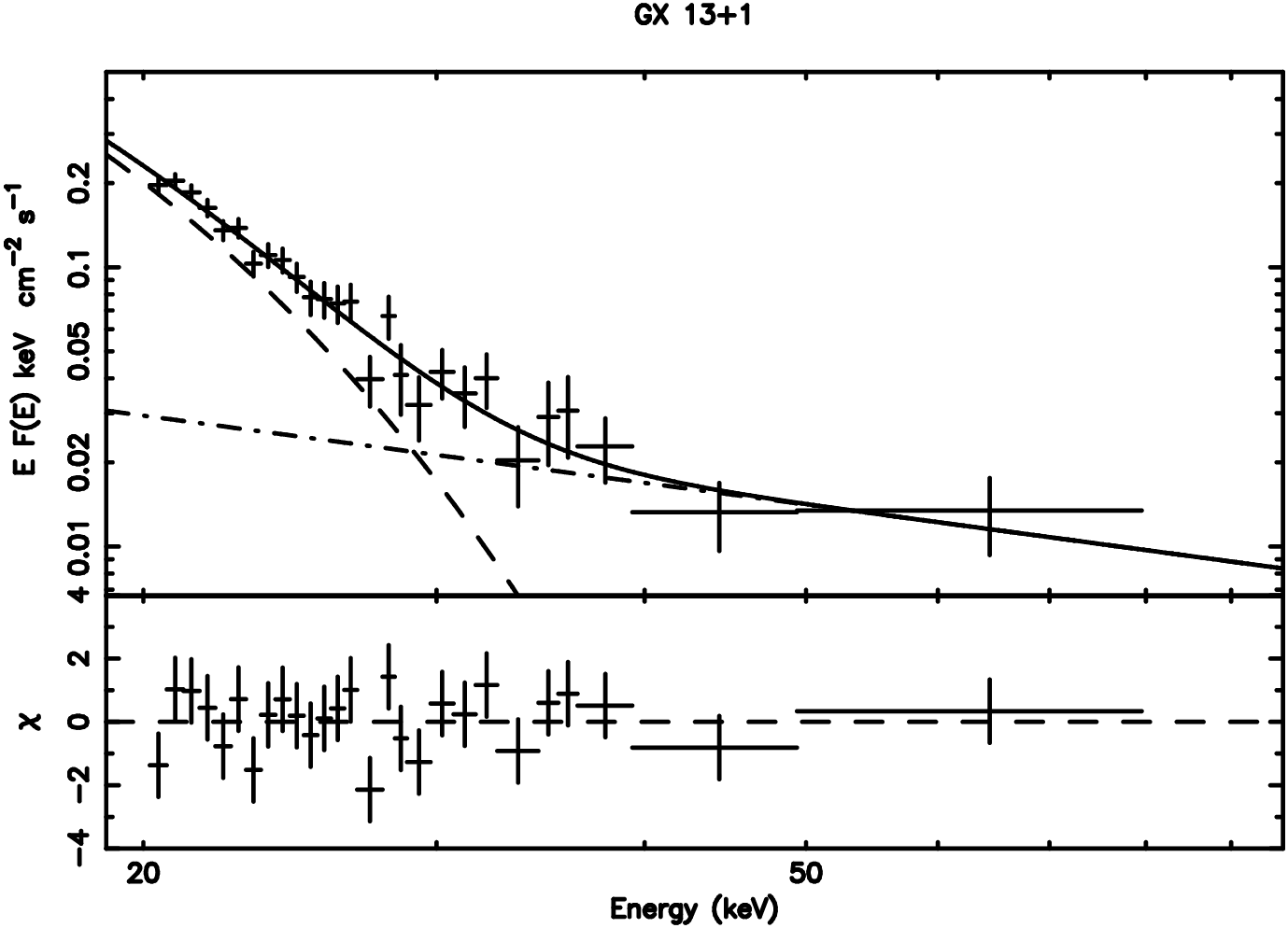}}

\caption[]{Atoll sources:  average IBIS/ISGRI spectra and best fit models shown in Table~\ref{tab:fit}.}
\label{fig:ASPE}
\end{figure*}

The fit we obtain for \textbf{GX~3$+$1} is not good ($\chi^{2}$/dof=46/24); however, as can be seen 
in Fig.~\ref{fig:ASPE},  
this is not due to a clear excess or systematics, more likely it is due to source spectral variability.
Our average spectrum parameters show that also GX~3$+$1 has a Z-like spectrum, 
similarly to what found by \cite{oosterbroek01} using \textit{BeppoSAX} data (optical depth $\tau$ about 6 
and Comptonising plasma temperature  kT$_{e}$ about 2.7\,keV).

\cite{iaria05} presented the 0.12-18\,keV spectrum of \textbf{GX~9$+$1} 
from a long \textit{BeppoSAX} observation.  The 0.12-18\,keV spectrum could be well fitted by 
a blackbody model plus Comptonised component, the latter with an electron temperature kT$_{e}$ of 2.3\,keV and 
optical depth $\tau$ of about 15. No firm conclusion could be drawn above 20\,keV due to 
source confusion issues. 
The \textit{INTEGRAL} observation above 
20\,keV does not suffer from source confusion and shows that the spectrum of GX~9$+$1
is very similar to the bright Atoll sources GX~9$+$9 and GX~3$+$1, i.e. very similar 
to the typical Z source spectrum. There is a hint of excess  in the spectrum of GX~9$+$1, Fig.~\ref{fig:ASPE}, 
but it is not significant with the current exposure time.

\subsubsection{The intermediate spectral state} \label{sec:hardtail}
A simple \comptt~model was not good enough to 
describe the data due to a systematic excess above $\sim$30\,keV for \mbox{Sco~X--1} ($\chi^{2}$/dof$>$60), 
\mbox{Cyg~X--2} (45/27), GX~17$+$2 (61/31), GX~5--1 (96/31), and  GX~13$+$1 (38/24). 
With the addition of 
a power-law, the fits improved significantly (see Table~\ref{tab:fit}) with the temperature of
 the Comptonising plasma, kT$_{e}$, 
stabilising itself around 2.5--3\,keV. 
Within this sample, transient hard tails had already 
been observed in the spectra of Sco~X--1 \citep{damico01} based on  \textit{RXTE} data, \mbox{GX~5--1} 
\citep{paizis05} based on \textit{INTEGRAL} data, 
\mbox{GX~17$+$2} \citep{farinelli05} and \mbox{Cyg~X--2} \citep{disalvo02b} 
based on \textit{BeppoSAX} data.
In our study, the seed photon temperature could not  be constrained and 
 was fixed to the value [1.0-1.3]\,keV for 
all the sources, with the exception of GX~340$+$0 for which it was fixed at 
0.93\,keV \citep{lavagetto04}, and GX~17$+$2 fixed at
0.6\,keV \citep{farinelli05}.

The spectrum 
of \textbf{GX~340$+$0} could be fitted with a simple
\comptt~model ($\chi^{2}$/dof = 32/29) with kT$_{seed}$ fixed at 0.93\,keV as found
by \cite{lavagetto04}, providing kT$_{e}$=$6.5^{+6.7}_{-2.1}$\,keV and $\tau$=$1.7^{+2.8}_{-1.2}$.
We note however, that if we freeze kT$_{e}$ to 3\,keV, as found in the \textit{BeppoSAX} broad-band 
fit (0.1-100\,keV) by \cite{lavagetto04}, around $\sim$40\,keV
the residuals systematically deviate from \comptt~and the fit is no longer
acceptable ($\chi^{2}$/dof=58/27).
In this case the addition of a power-law ($\Gamma$ frozen at 2.5, as
found in \textit{BeppoSAX} data) is required.

The Comptonising plasma parameters obtained in the fit of the spectrum of \textbf{Sco~X--1} 
show values typical of Z sources (kT$_{e}$ of a few keV and 
optical depth between 5-15). The excess left by this model is well fitted by a power-law 
with no measured cut-off up to 150\,keV. Such an extended 
 power-law
can be an indication 
for bulk motion (dynamical) Comptonisation  or non-thermal electron population given the 
 bad fit obtained with different {\it thermal} Comptonisation models.
See \cite{disalvo06} for a study of the spectral variability of Sco~X--1 with \textit{INTEGRAL}.

\cite{farinelli05}  used
the same model used here (\comptt~plus
power-law)  to describe the spectrum of \textbf{GX~17$+$2}, so a more quantitative comparison is possible. 
We fixed the power-law slope ($\Gamma = 2.7$), optical depth ($\tau = 11$) and seed photon temperature
($kT_{seed} = 0.6 \rm \, keV$) to what was found by \cite{farinelli05} and find a  similar 
electron gas temperature but a lower power-law normalization at 1\,keV 
(our $\sim$0.23 versus their $\sim$2.1).  This is compatible with a picture of 
a fairly constant soft component and transient hard tail component that is 
at maximum when observed in a pointed observation \citep{farinelli05} and
at a lower value when averaged on long-term observations (due to its transient nature).

Similarly, for \textbf{Cyg~X--2} the power-law slope could not be constrained and 
we used the average value found by \cite{disalvo02b} who also used the \comptt~plus power-law model.
The hard tail was the strongest in the horizontal branch and we used the photon index average value 
from the upper and lower horizontal branch ($\sim$2). The same was done for the seed 
photon temperature ($\sim$1\,keV) and optical depth ($\sim$11). Similarly to GX~17$+$2, 
we obtain an electron gas
 temperature close to what found by \cite{disalvo02b} and a power-law normalization 
 of $\sim$0.02 (to be compared to their $\sim$0.03). The difference in the power-law normalization is 
 much less than for the case of GX~17$+$2. This could be due to the fact that \mbox{Cyg~X--2} spends  more 
 time than GX~17$+$2 on the horizontal branch and so the  average spectrum of Cyg~X--2 is closer to the 
 horizontal branch spectrum than in the case of GX~17$+$2.
 
The hard X-ray spectrum of \textbf{GX~5--1} has always been difficult to study due to the presence of the
nearby black hole candidate GRS 1758--258 (40$^{\prime}$ apart). \cite{asai94} first detected a
possible hard tail in the spectrum of GX~5--1 although contamination from the nearby black 
hole could not be excluded. The first non-contaminated observation of a systematic hardening of the 
source spectrum in the horizontal branch  was done by \cite{paizis05} with \textit{INTEGRAL} 
data and the presence of the hard tail is confirmed here, based on a larger data set.

The addition of the power-law component in the spectrum of \textbf{GX~13$+$1} required some
parameters to be frozen. We fixed the seed photon temperature and 
optical depth to the values found by  \cite{homan04} (1 and 8\,keV, respectively).
The photon index was found to be steady around  2.8 (even with a cut-off \PL~plus \PL~model and 
bremsstrahlung plus \PL). Fixing the slope to 2.8, we obtained 
the Comptonising plasma 
temperature of about 2.75\,keV given in Table~\ref{tab:fit}.
The nature of GX~13$+$1 has always been a matter of debate: 
\cite{hasinger89} classified it as a bright Atoll source (hence our classification), 
although they noted that it showed properties which 
were closest to those seen in the Z sources; 
\cite{schulz89} grouped it with the high luminosity sources that  
  included the six sources that were later classified as Z sources;
\cite{homan98} discovered quasi periodic oscillations (QPOs) similar to the horizontal 
branch oscillations found in Z sources;
based on simultaneous radio and X-ray observations, \cite{homan04}
showed that the X-ray and radio luminosities of GX~13$+$1 
are actually in the range expected for Z sources unlike 
the Atoll sources GX~9$+$1, GX~9$+$9 and GX~3$+$1 that have similar X-ray luminosities
 but much lower radio ones. 
Our work shows that indeed  GX~13$+$1 behaves like a Z source not only due to its 
variability and radio properties but also from the 
\emph{spectral point of view} with an average spectrum very similar to 
 Z sources, i.e. thermal Comptonisation plus hard tail.

\subsubsection{The low/hard spectral state } \label{sec:PL}
The two remaining  Atoll sources, \mbox{H~1608--522} and \mbox{H~1705--44},  are the only ones of our
sample that can be well described by a single power-law. 
The burster \textbf{H~1608--52} has been detected beyond 100\,keV also in the past 
\citep[][with BATSE observations]{zhang96} when it was in a very bright state (about 100\,mCrab, 20--100\,keV)
with a spectral break or cut-off around 65\,keV. 

\textbf{H~1705--44} has been observed to show dramatic spectral changes in its X-ray spectrum
\citep{barret02}. 
 The 2--80\,keV could be well fitted by a blackbody plus Comptonisation 
model (\comptt) with a temperature of about 15\,keV i.e. the spectrum was not detected above 
100\,keV. In this study we detect the source up to about 150\,keV,
that, to the best of our knowledge, is a new result. The high sensitivity of IBIS/ISGRI 
combined with the long observation time (307\,ksecs) are essential ingredients for this new 
detection, similarly to what happened e.g. in the case of Aql~X--1 \citep{rodriguez06} and 4U~1850-087 
\citep{sidoli06}.

To have a feeling of the plasma temperature in \mbox{H~1608-522} and \mbox{H~1705-440}, we fitted their spectra 
with the Comptonisation model by \cite{poutanen96} (\compps~in \XSPEC~terminology) obtaining 
Comptonisation of soft seed photons ($\sim$2\,keV) by an optically thin ($\tau$$<$1)
hot plasma. 
We assumed a spherical geometry and a blackbody seed photon spectrum (Table~\ref{tab:fit}).

\subsection{Generic Comptonisation Model}

As we have mentioned above, in our analysis of X-ray spectra from NS LMXBs  
we found  similarities 
 with black hole LMXBs. The BHC and NS spectral shapes are generic: they  
consist of blackbody-like (BB) and power-law components. Sometimes one needs an exponential 
rollover in order to terminate the power-law component at high energies. The main difference is 
that NS spectra are usually softer for the same state.  Another difference in 
these spectra is that in the NS case there are two BB components (not one as in BHC spectra) 
 which can be related to the emission from the disc (BB colour temperature  about 1\,keV) 
 and NS surface (BB colour temperature about 2.5\,keV), (TS05).

\cite{titarchuk96,titarchuk97}, hereafter TMK96 and TMK97, introduced the Generic 
Comptonisation model (\BMC~model in \XSPEC) that takes into account the dynamical Comptonisation 
(converging inflow, expected in the vicinity of the central object)  along 
with thermal Comptonisation. It is worth noting that this model can be applied to fit the observed 
spectra of NS and BHC sources. The \BMC~model reads as:
\begin{equation}
F(E)=\frac{C_N}{1+A}(BB+A\times BB\ast G)
\label{bmc_spectrum}
\end{equation}
where $G(E,E_0)$ is the Green's function obtained as a solution of radiative diffusion equation in 
which both effects of thermal and dynamical Comptonisation are included. 
$BB$ stands for the injected blackbody-like spectrum, namely $BB\propto  B_{\nu}(T_{col})$,
$BB\ast G$ is a convolution of $BB$ with the (Comptonisation)  Green's function $G(E,E_0)$,
$C_N$ is the BB normalization coefficient (that depends on the compact object mass and  distance 
to the source). The factor  $1/(1+A)$ is the fraction of the seed photon radiation 
directly seen by the Earth observer, whereas the factor $A/(1+A)$ is the fraction of 
the seed photon radiation up-scattered by the Compton cloud.

 $G(E,E_0)$ is the response of the Compton medium (cloud) to the injection  of monochromatic 
 line with energy $E_0$.    TMK97 showed that $G(E,E_0)$ is a broken power-law with  blue wing 
index $\alpha$ and red wing index always more than $ \alpha+3$. 
Note that  $G(E,E_0)$ approaches Delta-function $\delta (E-E_0)$ for \mbox{$\alpha\gg1$}. 
High values of $\alpha$ are an indication of low efficiency of Comptonisation that  usually occurs 
when the plasma of  the Compton cloud is very close to equilibrium with the seed photon environment.  
The electron temperature of the Compton cloud differs by a factor of a few from the BB 
seed photon temperature.
We also have to note that a single BB spectrum can be reproduced 
either with $A\ll1$ (small coverage of the seed photon radiation by the Compton cloud) or with $A\gg1$ 
(large coverage) \emph{and} $\alpha\gg1$
(inefficient Comptonisation). 
The free parameters of the \BMC~model (apart from the normalization $C_N$) are the BB colour 
temperature, $kT_{bb}$, the spectral index $\alpha$  (photon index $\Gamma=\alpha+1$) and $\log A$.   

We can describe all the spectra studied in this paper 
   in terms of  bulk motion Comptonisation.
The results of the fits of our \textit{INTEGRAL} spectra 
with the \BMC~model are reported in Table~\ref{tab:bmc}. 
For \mbox{GX~340$+$0}, \mbox{GX~349$+$2}, \mbox{GX~9$+$9}  and \mbox{GX~3$+$1} 
only $kT_{bb}$ is well constrained.
This is not surprising given that in their spectra only the thermal bump is unambiguously detected. 
The best-fit colour temperatures are  about $2.5$\,keV using \comptt~or  \BB~models.
GX~17$+$2 and GX~13$+$1 have similar best-fit $\Gamma$-values, even if not well constrained, whereas 
the slope of GX 5-1 is steeper but much better constrained than in GX~17$+$2 and GX~13$+$1.
The upper value of $\Gamma$ reported for Cyg~X--2 (\mbox{$\lesssim$ $3$}) could be indicative  
of the presence of photon up-scattering (Comptonisation) of BB photons originated in the disc 
and NS surface. One can arrive to  similar conclusions for \mbox{GX~9$+$1} even if the effect seems 
to be less prominent there (\mbox{$\Gamma\lesssim$ $4$}).
For the two dim Atoll sources \mbox{H~1608--522} and \mbox{H~1705--440}   we find that 
$\Gamma$ is well constrained while there is no real information on $kT_{bb}$. This is not 
surprising given that in their spectra we do not see evidence of the thermal bump (described by 
the \BB~in the \BMC~model) and so it is impossible to infer either the BB colour-temperature or its relative 
contribution to the merging flux. 
In Sco X-1, the whole set of the best-fit parameters is well constrained, 
given that both thermal bump and the hard tail are detected with high statistical significance.

Since our analysis uses data starting from 20\,keV, careful interpretation should be given to the 
origin of the BB seed photons found in  Table~\ref{tab:bmc} (kT$_{bb}$$\sim2.5$\,keV).  A broad band analysis 
of  GX~17$+$2 and Cyg~X--2 spectra (Farinelli, private communication) seems to  suggest that, similarly to the BHC case, 
the seed photons for the \BMC~model are rather soft, with a temperature close to disk emission values,  
$\sim$ 0.4-0.6 keV \citep{barret00}. 
 A second Comptonisation component, \comptt, is also needed with $\sim$2.5\,keV seed photons coming most likely from the NS surface. 
 In our spectra the disk component is too soft to be detected  and the \BMC~model interpretes the $\sim$2.5\,keV seed photons 
as due to the emission from non-efficiently up-scattered seed photons within the bulk flow.
The energy sampling of our data does not allow us to produce a complete discussion
about the full geometry of the sources and global \BMC~model parameters, beyond the scope of this article,  
and this caveat should be kept in mind in all we discussed above that remains, in any case, qualitatively valid.


\section{Discussion}
We have analysed  IBIS/ISGRI available public data on a sample of twelve persistent 
LMXBs containing a neutron star.
We focused our study on the average spectral behaviour of the sources and 
classified them in terms of  spectral states.
As shown by the light-curves we presented, the sources have 
 some degree 
of variability, but the average spectrum is 
a representation of the sources above 20\,keV and enables the study of the global properties. 

\subsection{A scenario for spectral evolution in NS LMXBs}
We observe three main spectral states: 
a \emph{very soft} state  with source spectra that can be well described by a 
single thermal Comptonisation component,  
an \emph{intermediate} state (thermal Comptonisation plus 
power-law)
and a \emph{low/hard} state  (single power-law). 
We have  successfully studied these three spectral states in the 
 frame of the Generic Comptonisation Model (\BMC, TMK96).
We present  our scenario of the spectral evolution of NS LMXBs 
starting from the low/hard state. 

The {\it low/hard state} is characterized by a low mass accretion rate  
in the disc. In this case the gravitational energy release in the disc is much 
smaller than the one in the optically thin outer boundary of the corona (Compton cloud). 
The coronal outer boundary   is presumably related to the adjustment shock 
\citep{titarchuk04}. 
The corona completely covers the seed photon area (high $\log A$) 
and the emergent spectrum is a result of the  up-scattering (Comptonisation) 
of the seed photons in the corona. At the very low level of accretion rate, 
the bulk motion is extremely weak and the emergent spectrum can be interpreted as pure thermal 
Comptonisation with no bulk signature (GX~354$-$0 in Fig.~\ref{fig:comp}). 
The bulk motion effect increases with accretion rate and the power-law signature starts to be visible in the spectrum, 
the cut-off is moved at a higher energy than what expected from pure thermal Comptonisation 
(H~1608-522 in Fig.~\ref{fig:comp}).
In either case,
one cannot see  any trace of the seed photons 
in ISGRI (that give the BB bump). Open magnetic field lines are exposed to the observer 
 but the outflow is weak because of the low mass accretion rate (there is not enough radiation 
 in the disc to launch the wind) and the  system is radio-quiet. 
  
  In 
  the {\it intermediate state} the mass accretion rate increases with respect to  the
   low/hard state. It leads to high efficiency of the Comptonisation, particularly 
   bulk inflow Comptonisation that is seen as  an  extended hard tail in the spectrum. 
    Thermal Comptonisation becomes less efficient because the coronal plasma is cooled 
    down by the seed photons coming from the disc and NS surface. The corona consists 
    of a quasi-spherical component (related to the closed field lines and bulk motion inflow) 
    and of a cylindrical component (related to the open field lines and outflow, TS05). The vertical size of 
    the cylindrical configuration is suppressed as the mass accretion increases. 
    We start to  see the seed blackbody bumps (only the higher energy one in the IBIS/ISGRI
    range), because the corona is cooled down and 
    becomes more compact. 
      
    The {\it very soft} state is characterized by a high mass 
    accretion rate that is very close to the critical (Eddington) values. The emergent 
     spectrum  is a sum of two blackbody-like spectra, one is related to the Comptonisation 
     of the NS photons (visible in the IBIS/ISGRI range) and the other one is related to the
      Comptonisation of the disc photons.
     The seed photon and plasma temperatures differ by a factor of a few.  The electron 
     plasma and the photons in the Compton cloud are very close to equilibrium.   In this 
     state the corona is quasi-spherical, there is  no bulk motion and no radio emission. In fact,
      the radiation pressure caused by the strong emission from the NS surface stops the 
      bulk inflow and the high accretion rate   changes the
 configuration of the field lines  and the radio emission is quenched.
 
It is important to note that what plays the key role to suppress the bulk motion is the 
 \emph{local} radiation pressure, hence the \emph{local} accretion rate impinging
the bulk inflow: Sco~X-1 has a total isotropically-estimated luminosity that is  higher 
than e.g. GX~3$+$1, but we speculate that the inflow anisotropy is much higher in GX~3$+$1 
than in Sco~X-1, resulting in a higher local radiation pressure, i.e. bulk suppression.

\begin{table}[t]
\caption{Comparison of the \emph{average} radio (8.5\,GHz) and X-ray hard tail (40--100\,keV) flux  of the 
Z and bright Atoll sources studied in this paper. 
The radio data are taken from \cite{fender00} and 
\cite{migliari06}.  }
\label{tab:radio}
\begin{center}
\begin{tabular}{lccc}
\hline
Source &  Radio flux         & Hard tail flux    \\
       &           (mJy)                    &  10$^{-11}$\rm\,erg\,s$^{-1}$\,cm$^{-2}$ \\
\hline
\hline
Z sources &       &                           & \\
\hline

Sco~X--1  & $10 \pm 3$ & 18 \\
GX~17$+$2  & $1.0 \pm 0.3$ &1.9 \\
GX~349$+$2 & $0.6 \pm 0.3$ & $<$1.15 \\
Cyg~X--2  & $0.6 \pm 0.2$ & 1.7 \\
GX~5--1   & $1.3 \pm 0.3$ &  1.9\\
GX~340$+$0 & $0.6 \pm 0.3$ & 2\\
\hline 
Atolls & \\
\hline
GX~9$+$1  &  $<$0.2 & $<$0.75 \\
GX~9$+$9  & $<$0.2 & $<$1.72 \\ 
GX~3$+$1  &  $<$0.3  & $<$0.43  \\
GX~13$+$1  & $1.8 \pm 0.3$ & 1.74 \\
\hline
\end{tabular}
\end{center}

\end{table}
\begin{figure}
\centering
\includegraphics[width=1.0\linewidth]{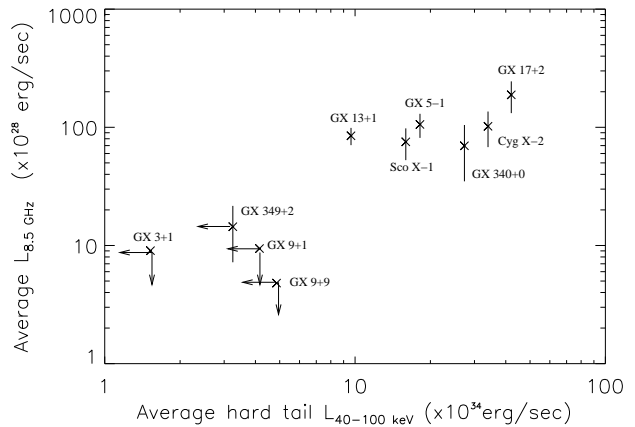}
	\caption{Average radio luminosity plotted against the average hard \mbox{X-ray} tail 
	(40--100\,keV) luminosity. The correlation is clearly visible also
	if we compare the source fluxes instead of the luminosities (see text).
\label{fig:corr}}
\end{figure}

\subsection{The radio emission - hard X-ray tail connection}
In the scenario proposed in the previous Section there is a clear connection 
between the X-ray spectral states and the radio emission: low/hard states are associated to 
weak radio emission, the radio emission increases in the intermediate state and then 
is quenched in the very soft state. 

This trend is clearly met in the observations. We  can imagine a continuous "accretion line" increasing
 from Atoll sources (island state to banana) moving 
to the  Z sources (horizontal, normal and flaring branch).
Only a few low-dim Atoll sources (typical low/hard state) 
have been detected in the radio band because of their low radio luminosity \citep[][]{fender00, migliari06}.
Z sources in their horizontal branch have a spectrum that corresponds to the intermediate state we defined here 
and indeed a clear radio emission has been reported in this state \citep[][and references therein]{fender00, disalvo02}. 
The normal and flaring branch of Z sources corresponds to the very soft state defined here where no 
radio emission is expected and in fact Z sources in these branches have a severely reduced radio emission, if any.


 Z sources are much brighter radio emitters than  Atolls and GX~13$+$1 has an important 
radio emission, thus
 behaves like a Z \citep[][]{fender00, migliari06}.
In average, unlike bright Atolls, Z sources have a hard tail besides the thermal Comptonisation by
an optically thick plasma of $\sim$3\,keV   
\citep[][and this work]{disalvo02};  GX~13$+$1 has a hard tail (this work), thus again
 behaves like a Z.
We basically see that in GX~13$+$1, the link radio emission - hard X-ray tail (both "against" 
its originally declared 
 nature to be an Atoll source) reveals itself in a solid way.

Triggered by the GX~13$+$1 case, we compared the average radio emission of all the sources 
of our sample with the  hard X-ray tail that appears on top of the $\sim$3\,keV Comptonisation 
spectrum.
We note that the hard X-rays from  the  two soft X-ray dim Atoll sources, \mbox{H~1608--522} and \mbox{H~1705--440}, 
are mainly coming from Comptonisation of soft photons in a hot corona (i.e. 
the overall spectrum is hard, kT$_{e}$ $>$ 40\,keV, see Table~\ref{tab:fit}) and since we 
cannot disentangle the dominating thermal Comptonisation component from the dynamical (bulk) one,
we do not include these two sources in the radio - hard tail study.
Table~\ref{tab:radio}  reports the average radio flux as derived by 
 \cite{fender00} and \cite{migliari06} and the 40--100\,keV flux of the hard tails detected in this work. 
When a hard tail is detected, its 40--100\,keV flux is computed after fixing the \comptt~normalization to zero.
For the sources where a hard tail was not detected, we computed 
 3$\sigma$ upper limits (3$\times$$error$ in the 40--100\,keV range), assuming that all
 the flux in the 40--100\,keV comes from the hard tail. Figure~\ref{fig:corr}
 is visualising the relation between the average radio and 
  hard tail \mbox{(40--100\,keV)} \emph{luminosities} computed using the distances in  Table~\ref{tab:table1}. 
The correlation between the two is quite striking (Spearman rank order significance of 99.6\%) and the
 upper limits are consistent with the trend. We note that we obtain a very good correlation 
 (Spearman rank order significance 99.2\%) also using the radio and hard tail \emph{fluxes} instead of the 
 luminosities. We chose to plot the luminosity correlation in order to show the intrinsic
 properties of the sources.

In Fig.~\ref{fig:corr} we have plotted the average radio and X-ray hard tail luminosities: 
the radio luminosities are the superposition of optically thick emission (compact jet) 
and optically thin flaring activity. The X-ray hard tail luminosities are the average of all hard tail states 
(from maximum to minimum strength). We are aware that these correlations are based on data that are not 
taken simultaneously, but 
they are all based on average fluxes  
that we can  consider a good representation of the physics involved in these objects.
For the sources where a hard tail has been observed, we compare the radio emission with what is \emph{left} 
in the X-ray spectra  of the sources once we \emph{remove}  
the dominant Comptonisation component. 
In this respect, our detection of a hard tail in the spectrum of GX~13$+$1 is important:  GX~13$+$1 has a 
similar X-ray flux (and luminosity) as the 
remaining bright Atoll sources (GX~3$+$1, GX~9$+$1, GX~9$+$9, see Table~\ref{tab:table1}) but a much higher 
 average radio and hard tail emission (Table~\ref{tab:radio}), 
re-confirming the radio  hard X-ray tail connection.

In the presence of a cut-off, the hard tail can be explained 
via Comptonisation of soft seed photons in  the jet and/or in  the corona.  
For the case where no cut-off is detected,
many models have been proposed: Comptonisation by a hybrid (thermal non-thermal) corona \citep{coppi99},  
synchrotron emission from the electrons of the jet \citep{markoff05}
or bulk motion inflow (dynamical) Comptonisation (TMK96, used in this work).

The real test of any of these models can be done using the (variability) analysis of the power density
 spectrum (PDS)  of the hard tail emission. 
 The characteristic (break and QPO) frequencies of PDS  do determine 
the geometric size  of the configuration  where the hard tail emission is formed.
 TS05   made this type of analysis  for 4U~1728--34  and  found evidence that  the hard 
 tails are  formed in the compact Compton 
 cloud with geometry changing from cylindrical-like in the low/hard state to the
  quasi-spherical one in the high/soft state.
Note that the X-ray-radio correlation along  with the QPO-radio correlation is
well established in 4U~1728--34. TS05 presented  an explanation of these correlations 
 in the framework of an oscillation model  using the observed correlations 
   of QPO low frequencies and their ratio.

The correlation we find between the radio and hard X-ray tail emission  suggests 
that the  hard tail formation area and the source of 
 energetic electrons, ultimately causing  
  the radio  emission,  are closely connected.
   The most 
 probable site of this configuration  is the NS magnetosphere.  
It can be suggested that the open magnetic field lines of the NS
 magnetosphere are the base of the jet seen in the radio emission.
 An increasing accretion rate leads to a more efficient radio
 emission (low/hard state to intermediate state) up to a point where the
 extremely high accretion rate (very soft state)  changes the
 configuration of the field lines  and the radio
 emission is quenched.

\begin{acknowledgements}
The authors thank A. Segreto and C. Ferrigno for the useful cross checks in the 
analysis phase. AP thanks M.~R. Gaber, B.~E. O'Neel and G.~A. Wendt III for
their precious contribution to the development of post-processing tools.
AP, AB and SM acknowledge the Italian Space Agency financial and programmatic support
via contract I/R/046/04.

\end{acknowledgements}

\bibliographystyle{aa}
\bibliography{biblio}

\end{document}